\documentclass[preprint,secnumarabic,amssymb,amsmath,footinbib,tightenlines,apl,prb,showkeys,longbibliography]{revtex4-1}

\usepackage{amsmath,amssymb,amsfonts,textcomp,amsthm,mathtools}
\usepackage{gensymb}
\usepackage{setspace}  
\usepackage{marvosym}
\usepackage{psfrag}
\usepackage{subfigure}
\usepackage[dvips]{graphicx}
\usepackage{subfigure}
\usepackage{natbib}
\DeclareGraphicsExtensions{.eps}
\usepackage{color}
\usepackage{eulervm}
\usepackage{capt-of}
\usepackage{epstopdf}
\usepackage{array}
\usepackage{multirow}
\usepackage{ mathrsfs }
\usepackage{morefloats}
\usepackage{times}
\usepackage{marginnote}

\begin{document}

\title{Instabilities and transition in cooled-wall hypersonic boundary layers}

\author{S. Unnikrishnan}
\affiliation{Mechanical Engineering, Florida State University, Tallahassee, Florida 32310} 

\author{Datta V. Gaitonde}
\affiliation{Mechanical and Aerospace Engineering, The Ohio State University, Columbus, OH 43210} 

%\marginnote{R1C1}[.01cm]

\date{\today}


\begin{abstract}
Wall cooling has substantial qualitative and quantitative effects on the development of instabilities and subsequent transition processes in hypersonic boundary layers (HBLs).
A sequence of linear stability theory, two-dimensional and non-linear three-dimensional direct numerical simulations is used to analyze Mach~6 boundary layers, with wall temperatures ranging from 
near-adiabatic to highly cooled conditions, where the second-mode instability is accompanied by radiation of energy.
Decomposition of linear stability modes into their fluid-thermodynamic (acoustic, vortical and thermal) components shows that this radiation comprises both acoustic as well as vortical waves.
Furthermore, in these cases, 2D simulations show that the conventional ``trapped'' nature of second-mode instability is ruptured.
A quantitative analysis indicates that although the energy efflux of both acoustic and vortical components increases with wall-cooling, the destabilization effect is much stronger and no significant abatement of pressure perturbations is realized.
The direct impact of these mechanisms on the transition process itself is examined with high-fidelity simulations of three-dimensional second-mode wavepacket propagation. 
In the near-adiabatic HBL, the wavepacket remains trapped within the boundary layer and attenuates outside the region of linear instability. 
However, wavepackets in the cooled-wall HBLs amplify and display nonlinear distortion, and transition more rapidly.
The structure of the wavepacket also displays different behavior; moderately-cooled walls show bifurcation into a leading turbulent head region and a trailing harmonic region, while highly-cooled wall cases display lower convection speeds and significant wavepacket elongation, with intermittent spurts of turbulence in the wake of the head region. 
This elongation effect is associated with a weakening of the lateral jet mechanism due to the breakdown of spanwise coherent structures. 
These features have a direct impact on wall loading, including skin friction and heat transfer.
In moderately cooled-walls, the spatially-localized wall loading is similar to those in near-adiabatic walls, with dominant impact due to coherent structures in the leading turbulent head region. 
In highly-cooled walls, the elongated near-wall streaks in the wake region of the wavepacket result in more than twice as large levels of skin friction and heat transfer over a sustained period of time.
\end{abstract}

\maketitle

\section{Introduction}\label{sec_intro}
The problem of managing drag and heat transfer changes associated with transition to turbulence in hypersonic boundary layers is predicated on a detailed understanding of the relevant instability mechanisms.
Hypersonic boundary layers (HBLs) harbor multiple instability modes, and display several possible routes to transition \citep{reshotko1976boundary}. 
The broad range of parametric sensitivity of HBLs to perturbation evolution introduces significant challenges. 
In addition to factors such as freestream conditions and geometrical features, a key parameter governing instability growth in HBLs is the wall thermal condition. 
The most unstable Mack mode (or second-mode) instability in HBLs \citep{mack1975linear,mack1984boundary} is particularly sensitive to the wall temperature. 
Cooling destabilizes this mode, making the HBL more susceptible to transition \citep{stetson1992hypersonic,mack1975linear}. 
In addition, the origin of this unstable second-mode shifts from the slow acoustic spectrum to the fast acoustic spectrum \citep{fedorov2001prehistory}. 
Cold-wall conditions are especially relevant when interpreting results from high-enthalpy experimental facilities, where the thermal inertia of test surfaces maintains wall-temperatures much lower than the recovery temperature. 

An important feature of the second-mode in the context of cold walls is the identity and phase-speed variation of the most unstable mode. 
Under certain conditions this mode could travel at supersonic speeds with respect to the freestream in outer boundary layer regions, resulting in radiation of waves into the freestream, and oscillatory behavior in the corresponding eigenfunction. 
These observations have been confirmed through prior theoretical and numerical studies \citep{mack1990inviscid,bitter2015stability,chuvakhov2016spontaneous}. 
\citet{knisely2019sound1} provide detailed schematics and describe the wall-normal profiles in the non-radiating and radiating modes. 

Recent efforts have also made progress on understanding changes in inter-modal dynamics as the wall is cooled.
\citet{knisely2019sound2} used direct numerical simulations (DNSs) to identify resonant-type interactions between the slow and fast discrete modes with the slow continuous acoustic spectra. 
This highlights a complex set of dynamics contributing to the genesis of phenomena, such as the supersonic mode, that strongly motivate the augmentation of theoretical approaches with  high-fidelity numerical simulations.
A particularly interesting observation, with significant potential practical implication, emerges from the
DNSs of \citet{chuvakhov2016spontaneous}, who report that radiation by supersonic instability waves arising under cooled-wall conditions redistributes energy from higher to lower frequencies. 
This may constitute a mechanism to mitigate instability growth, and eventually transition, in HBLs. 

The generation of unstable modes with supersonic phase speeds is related to the influence of wall-cooling on the mean-flow features. 
In general, cooling increases gradients in the boundary layer, which can affect perturbation evolution past nonlinear saturation and alter the associated spatio-temporal scales. 
The enhanced gradients, along with baroclinicity, can also strengthen the generation of vorticity \citep{shadloo2017laminar}. 
DNSs by \citet{salemi2018synchronization} reveal that a prominent signature of linear instability mechanisms persists even in the nonlinear evolution phase of 3D wavepackets on a wall-cooled hypersonic cone. 
\citet{huang2019assessment} showed that even under turbulent conditions, wall-cooling introduces more deviations from Boussinesq and Reynolds analogy assumptions. 
This resulted in erroneous predictions of Reynolds normal stresses and turbulent transverse heat fluxes, with implications for turbulence modeling of HBLs. 

The present work builds on the above understanding of the effect of wall-cooling on HBL transition by considering a spectrum of increasingly higher-fidelity techniques on a broad range of wall temperatures ranging from near-adiabatic to highly-cooled conditions.  
The primary objectives may be succinctly stated as follows:
\begin{enumerate}
\item Employ linear stability theory to characterize the changes to the second-mode across the different wall temperatures, including changes in identity, phase-speed and fluid-thermodynamic content that ultimately yield acoustic radiation. 
\item Use two-dimensional (2D) DNSs to document the changes in the ``trapped'' nature \citep{fedorov2011transition} of the second-mode instability with wall cooling, and to subsequently quantify the efflux of acoustic and vortical energy from the HBL due to supersonic mode radiation.
\item Assess, through three-dimensional (3D) DNSs, the impact of supersonic modes on the breakdown process, and identify the primary features of turbulence development in wall-cooled HBLs.
\end{enumerate}
To place the results in the proper context in the available literature, the flowfield studied corresponds to a freestream Mach number ($M_\infty$) of $6$ \citep{chuvakhov2016spontaneous}, and wall to freestream temperature ratios ranging from $7$ to $0.1$ (\S~\ref{sec_fanz}). 

% % ??? Begin integrate

% \textbf{??? DVG will integrate on second reading. A systematic quantification of the energy efflux from cooled HBLs could provide further insights into the efficacy of this mechanism. 
% Although the origin and characteristic features of instability modes are illustrated in the above references, the impact on  inter-modal interactions and effect on the downstream transition has  not been fully explored.???}

% % ??? End integrate

The dynamics of HBL stability modes has been well established for the adiabatic (hot) wall case.
Wave-speed synchronization of different eigenmodes of the linear stability theory (LST) equations, has been associated with critical changes in the dynamics of prominent instabilities.
For consistency, we adopt the convention in \citet{fedorov2011high}, which identifies the two discrete instability modes  relevant to HBLs as mode~S and mode~F. 
These discrete modes originate from the slow and fast acoustic wave speed limits, $(1-1/M_\infty)$ and  $(1+1/M_\infty)$, respectively.
The synchronization of mode~F and the continuous vortical/entropic spectra is associated with enhanced sensitivity of this instability to freestream turbulence or hotspots \citep{fedorov2003initial}. 
Mode~F synchronization with mode~S is followed by the latter becoming unstable for the second time (second-mode) in adiabatic and warm walls \citep{ma2003receptivity1}. 
In sufficiently cooled walls however, mode~F becomes unstable following this synchronization \citep{fedorov2001prehistory}, and it eventually synchronizes with the slow continuous acoustic spectra.
The previously noted supersonic radiating modes appear beyond this point.

In \S~\ref{sec_vreisp}, linear theory is first utilized to quantify these observations in the cold walls under consideration in this study. 
Key aspects, including changes in the identity of the most unstable mode (F or S) with wall cooling, are identified and the corresponding effects on the eigenfunctions are elaborated.
In order to better understand the observed radiation in cooled-wall layers, these stability modes are connected to their physical nature by decomposing their content into  fluid-thermodynamic (FT) or Kov{\'a}sznay-like components, using the procedures outlined in \citet{doak1989momentum,unnikrishnan2019interactions}.
For reasons summarized in context below, the FT components, \textit{i.e.,} fluctutations having pure vortical, acoustic and thermal character, are extracted using $\rho \boldsymbol{u}$, where $\rho$ is the density and $\boldsymbol{u}$ is the velocity vector. 
%This mapping into acoustic, vortical and thermal components facilitates a clearer understanding of the radiating and non-radiating characteristics of cooled-wall HBLs.
This procedure complements the mathematical insight obtained from LST and provides novel insights into the nature of waves radiated from the supersonic mode, such as whether, for example the radiation is purely acoustic in nature \citep{salemi2018synchronization}.

The influence of radiation on the initial receptivity and linear growth processes cannot be fully portrayed by the linear stability modes.
A natural next step, discussed in \S~\ref{sec_lnpran}, is to examine the growth of linear harmonic perturbations in the 2D boundary layer.
Such simulations recreate the expected behavior of the second-mode instability under various cold-wall conditions. 
The splitting into FT components is repeated in this case;  the feature that the split components of $\rho \boldsymbol{u}$, represent mass fluxes having direction accrues an immediate benefit that it can be used to examine how acoustic flux lines differ between non-radiating and radiating boundary layers.
Specifically, the anticipated change from a compact trapped acoustic component at higher wall temperatures to a non-compact radiating one when the wall is cooled, is examined in detail.

The FT content, while insightful, does not directly inform the question of how radiating modes influence transition.
One approach is to quantify such effects through the resulting energy efflux from the HBL.
%??? See Comment
A highly effective way to quantify this energy loss is to examine the flux of the total fluctuating enthalpy (TFE), $H'$ where $H$ is the total enthalpy ($H = c_p T +{\boldsymbol{u.u}}/{2}$, with $c_p$ and $\boldsymbol{u}$ being the specific heat at constant pressure and the velocity vector, respectively).
The method has previously been successfully employed to examine the sources and sinks of near-wall acoustic energy in the second-mode for adiabatic walls \citep{unnikrishnan2019interactions}.
In the present work, the technique is applied to the outer part of the boundary layer for each wall condition, to provide a uniform assessment of energy losses associated with boundary layer radiation.
%This is achieved by performing linear perturbation analysis in \S~\ref{sec_lnpran}, through a high-order numerical approach. 
%The TFE budget facilitates a quantitative assessment of the energy efflux into the freestream with wall temperature change.
The discussion then assimilates the effect of radiation on transition dynamics, and the potential for transition mitigation.

A realistic understanding of the breakdown process following non-linear saturation requires a fully three-dimensional analysis, which can yield crucial insights on changes in fundamental mechanisms across broad parameter ranges where transition sensitivity is observed.
%High-fidelity simulations addressing the problem of highly-cooled boundary layers are relatively scarce, however.
For example, using DNSs, \citet{jocksch2008growth} observed streamwise elongated tails and slower convective speeds of turbulent spots in a moderately cooled Mach~$5$ boundary layer, compared to those in HBLs  with a near-adiabatic wall temperature. 
At similar levels of cooling, \citet{redford2012numerical} report spanwise coherent structures that extend into the calmed region behind the turbulent spot in a Mach~$6$ boundary layer. 
% ??? See comment
These structures are attributed to the destabilized second-mode instabilities that coexist with the turbulence in the spot. 
In \S~\ref{sec_trnchbl}, 3D DNSs are performed on a range of wall temperatures; the boundary layers are excited using wavepackets, whose features are informed by the linear theory and perturbation analyses discussed above.  
In addition to changes in the physical characteristics of the wavepacket structure as it propagates downstream at different wall temperatures, several pertinent trends are identified in the evolution of nonlinearities and breakdown mechanisms.
The consequences on near-wall pressure signature and wall loading are emphasized.
%, and changes in energy effluxes for this 3D wavepacket-based situation are also delineated.
A detailed spectral analysis of disturbance evolution is presented to further quantify variations in the spatio-temporal features of turbulence generation.   
The consequent considerations for transition-prediction model development (see \textit{e.g.,} \citet{park2009wall}) are also summarized.

%??? DVG COMMENT
The flowfield conditions employed, meshes and algorithmic details, and aspects of the basic state at different wall temperatures are discussed in \S~\ref{sec_fanz}.
The influence of wall cooling on the eigenspectra obtained from the linear stability equations, and their decomposition into fluid-thermodynamic components are presented in \S~\ref{sec_vreisp}.
2D simulation results using LST results to inform initial and boundary conditions are probed in \S~\ref{sec_lnpran} to understand the pathways by which the radiating phenomena are established, and to quantify the energy efflux associated with the supersonic mode.
Finally, in \S~\ref{sec_trnchbl}, features of 3D wavepacket propagation under different wall thermal conditions are described, including focus on the establishment of skin friction and heat transfer.

% ASSIMILATE - START
%Informed by the linear theory and perturbation analyses, 3D DNSs are performed, where the cooled-HBLs are excited using wavepackets composed of supersonic modes.  
%the dynamics of supersonic modes in cooled HBLs, to evaluate possibilities of transition mitigation. 
% ASSIMILATE - END

%\section{Flowfields analyzed}\label{sec_fanz}
\section{Basic state features with wall cooling}\label{sec_fanz}
A flat plate boundary layer subject to a range of wall thermal conditions is considered at a freestream Mach number, ${M_\infty}=6$.
To establish a reference with existing studies, the flow conditions are chosen to be similar to those described for a sharp-edged plate in \citet{chuvakhov2016spontaneous}. 
The Reynolds number based on freestream parameters, $Re=\rho_\infty^{*}U_\infty^{*}L^{*}/\mu_\infty^{*}$, is $1\times10^6$,  where $(.)^*$ indicates a dimensional quantity,
$\rho_\infty^{*}$, $U_\infty^{*}$ and $\mu_\infty^{*}$ are the freestream density, velocity and dynamic viscosity, respectively, and $L^{*}$ is a reference length scale. 
A perfect gas model is adopted, with Prandtl number, $Pr=0.72$, and ratio of specific heats, $\gamma = 1.4$.
The temperature dependence of viscosity is modeled using the Sutherland law.
The Cartesian coordinates, $(x,y,z)$, represent streamwise, wall-normal and spanwise directions, respectively, with $x=0$ denoting the leading edge of the plate. 
$(u,v,w)$ are the velocity components in the Cartesian coordinate system, and $p$, $\rho$ and $T$ denotes pressure, density and temperature, respectively. 
In the discussion below, an over-bar, $\overline{(.)}$, represents a time-averaged quantity, and a prime, $(.)'$, denotes a perturbation component.

The impact of wall-cooling on instability properties relevant to transition in the HBL is evaluated with LST, 2D and 3D DNSs at eight wall to freestream temperature ratios, $T_W=T_W^{*}/T_\infty^{*}$, summarized in the computational matrix of table~\ref{tab:casdcr}. 
%The methods utilized include local linear stability theory (LST), as well as two-dimensional (2D) and three-dimensional (3D) DNSs. 
% simulates eight different values of $T_W$.  
\begin{table}
  \begin{center}
  \def~{\hphantom{0}}
  \begin{tabular}{lcc}
      $T_W$  & Analyses \\[3pt]
       $7$     & 1D-LST, 2D-DNS, 3D-DNS  \\
       $5$     & 1D-LST, 2D-DNS          \\
       $3$     & 1D-LST, 2D-DNS, 3D-DNS  \\
       $1$     & 1D-LST, 2D-DNS, 3D-DNS  \\
       $0.7$   & 1D-LST, 2D-DNS          \\
       $0.5$   & 1D-LST, 2D-DNS, 3D-DNS  \\
       $0.3$   & 1D-LST, 2D-DNS, 3D-DNS  \\
       $0.1$   & 1D-LST, 2D-DNS, 3D-DNS  \\
  \end{tabular}
  \caption{Cases studied with varying degree of wall-cooling}
  \label{tab:casdcr}
  \end{center}
\end{table}
$T_W=7$ corresponds to the near-adiabatic wall temperature for this freestream flow, while the other choices result in moderately \citep{bitter2015stability} and highly \citep{wright1977flight,chuvakhov2016spontaneous} cooled conditions discussed in the literature. 
The lowest values considered aid in identifying the most impactful trends in the evolution of stability and transition mechanisms. 
%Below, we first summarize the variations induced in the laminar flow due to wall-cooling, before addressing the associated dynamics in the following sections.

\begin{figure}
\centering
\setlength\fboxsep{0pt}
\setlength\fboxrule{0pt}
\fbox{\includegraphics[width=5.0in]{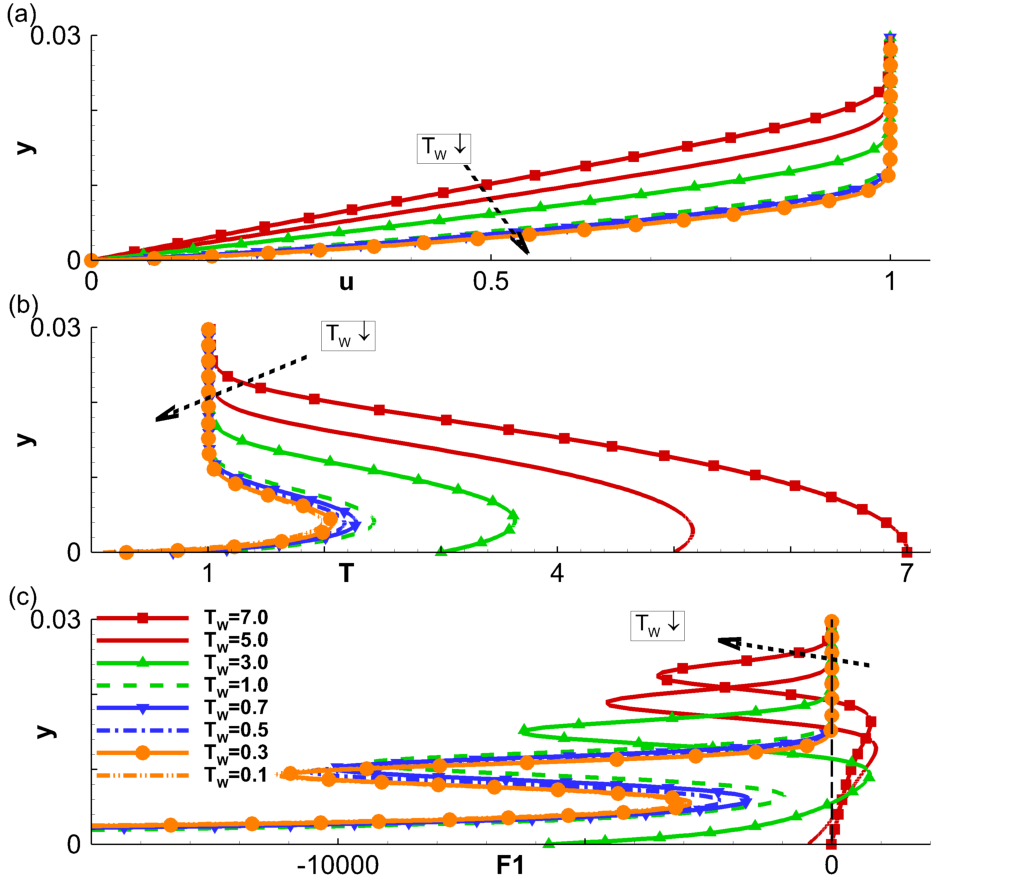}}
\caption{Effect of wall-cooling shown using similarity profiles of (a) streamwise velocity, (b) temperature, and (c) function, F1. 
The arrows indicate the trends in the profiles with decreasing values of wall temperature. 
The profiles are plotted at a streamwise location, $x=2$. 
The vertical dashed line in (c) marks $F1=0$.}
\label{figmnprfl}
\end{figure}

The effect of wall-cooling on the laminar basic state is summarized in figure~\ref{figmnprfl} using similarity profiles obtained from the Levy-Lees similarity solution \citep{anderson2000hypersonic}. 
The results are plotted at a representative streamwise location, $x=2$, in the non-dimensional physical coordinate system. 
Figures~\ref{figmnprfl}(a) and (b) display the velocity and temperature profiles, respectively.
Figure~\ref{figmnprfl}(c) shows the function $F1=\frac{\partial}{\partial y}\left(\frac{1}{{T}}\frac{\partial {u}}{\partial y}\right)$ to identify generalized inflection points (GIPs) in the laminar flow, which are an indication of inviscid instability. 
As evident in figures~\ref{figmnprfl}(a) and (b), the hydrodynamic and thermal boundary layers progressively become thinner as the surface is cooled and the flow density increases. 
This leads to higher near-wall gradients and correspondingly higher wall-loading. 
These trends are also manifested in the profiles of $F1$; specifically, the $y$-location of the peak value of the gradient appears closer to the wall with increased cooling. 
Prior to breakdown, both first and second-mode instabilities usually exhibit maximum amplitudes in the vicinity of this high-gradient region; thus transition dynamics is enhanced nearer the wall for colder surface conditions.

Zero-crossings of $F1$ indicate the existence of GIPs.
For the near-adiabatic wall, $T_W=7$, a single GIP is observed at $y\sim0.02$.  
With moderate cooling, a second GIP appears in the inner boundary layer, as seen for $T_W=5$ and $T_W=3$. 
With further wall cooling, the upper and lower GIPs arise closer to each other, and eventually, effectively cancel each other.
For the cases reported here, profiles for $T_W=1$ and lower do not exhibit a GIP. 
\citet{zhang2016effect} note that the absence of a GIP does not prevent the laminar profile from harboring inviscid instabilities. 
For cold walls, a region could exist in the boundary layer where the absolute relative Mach number, $|M_R|$, of an instability wave is locally supersonic with respect to the mean flow. 
Here $M_R=(u-c)/a$, where $c$ is the phase speed of the instability and $a$ is the local speed of sound. 
$|M_R|>1$ corresponds to an inviscid instability of the Rayleigh equation \citep{lees1946investigation,zhang2016effect}; in this manner, highly-cooled boundary layers can support second-mode instabilities. 

%\section{Variations in eigenspectra}\label{sec_vreisp}
\section{Eigenspectra and their fluid-thermodynamic content}\label{sec_vreisp}
Laminar profile variations resulting from wall-cooling profoundly affect linear instability mechanisms in boundary layers, as noted by several researchers (see {\em e.g.} the discussion by \citet{knisely2019sound1}). 
Prior to evaluating its impact on transition pathways, it is instructive to characterize the key mechanisms using a simplified temporal stability analysis.
%for the cases listed in table \ref{tab:casdcr}.
For this, we first obtain the local eigenspectra for various wall temperatures to track the evolution of the relevant unstable mode, and subsequently analyze its fluid-thermodynamic composition.

\subsection{Linear stability results}\label{sec_lsrls}
\begin{figure}
\centering
\setlength\fboxsep{0pt}
\setlength\fboxrule{0pt}
\fbox{\includegraphics[width=5.0in]{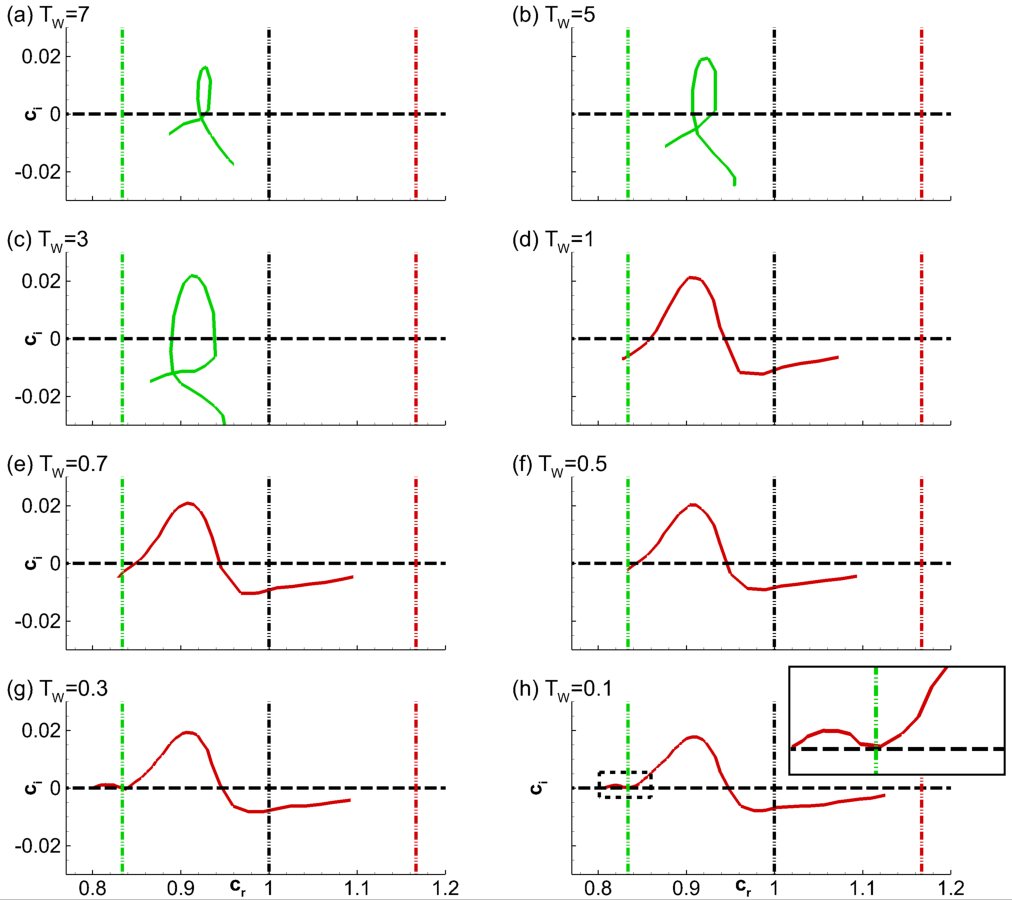}}
\caption{Loci of the unstable discrete mode obtained for the wave with $\lambda=0.025$.
Green and red loci have their origins near the slow and fast continuous acoustic spectra, respectively.
The horizontal dashed lines mark $c_i=0$.
The vertical dashed-dotted lines on the left, center and right mark the wave-speeds, $(1-1/M_\infty)$, $1$ and $(1+1/M_\infty)$, respectively. 
The inset in (h) magnifies the region marked within the dashed rectangle.}
\label{figegloc}
\end{figure}

The loci of unstable discrete modes are shown in figure~\ref{figegloc} for the different wall temperature conditions.  
%The plots represent the loci of the unstable discrete mode for various wall temperatures, as indicated. 
The horizontal and vertical axes are the real (speed, $c_r$) and imaginary (amplification rate, $c_i$) components of the phase speed of the instability wave. 
The horizontal dashed line marks the neutral limit while the vertical lines on the left, center and right mark the wave-speeds $(1-1/M_\infty)$, $1$ and $(1+1/M_\infty)$, respectively corresponding to the slow acoustic, entropic/vortical and fast acoustic waves in the freestream.
These results are obtained for a wave with streamwise wavelength, $\lambda=0.025$, at various streamwise locations in the range, $1\le x \le 7$. 
For the near-adiabatic wall, $T_W=7$, mode~S (shown green in these figures) becomes unstable as seen in figure~\ref{figegloc}(a), resulting in second-mode instability. 
Mode~F remains damped throughout this streamwise extent, and is thus not shown. 
This feature remains similar for moderate wall cooling, as seen for $T_W=5$ and $T_W=3$. 

For the next case analyzed, $T_W=1$, mode~S is stabilized, and second-mode instability now results from the positive growth rate of mode~F  \citep{fedorov2001prehistory}, whose locus, shown in figure~\ref{figegloc}(d) is plotted in red.
%The locus of  has its origin near $c_r=(1+1/M_\infty)$ s therefore plotted red. 
All lower wall temperature values, $T_W \le 1$ show similar behavior (figures~\ref{figegloc}(d)-(h)).
Quantitative trends are clearly discernible:
as the wall is  progressively cooled, the instability of mode~F extends to lower phase speeds, indicating that lower frequencies are appended into the unstable range. 
Figures~\ref{figegloc}(f)-(h) further show that for the wave parameters considered, a near-neutral mode~F interacts with the slow continuous acoustic spectrum. 
More interesting, for $T_W=0.3$ and $T_W=0.1$, a second zone of instability is present in the domain $c_r < (1-1/M_\infty)$, following a ``kink'' near this speed limit, consistent with the results of \citet{bitter2015stability}. 
This region of mode~F instability is highlighted in figure~\ref{figegloc}(h) using an inset. 
The presence of weakly damped or unstable mode~F at phase speeds lower than $(1-1/M_\infty)$ has been observed to result in radiation into the freestream. 

\begin{figure}
\centering
\setlength\fboxsep{0pt}
\setlength\fboxrule{0pt}
\fbox{\includegraphics[width=5.0in]{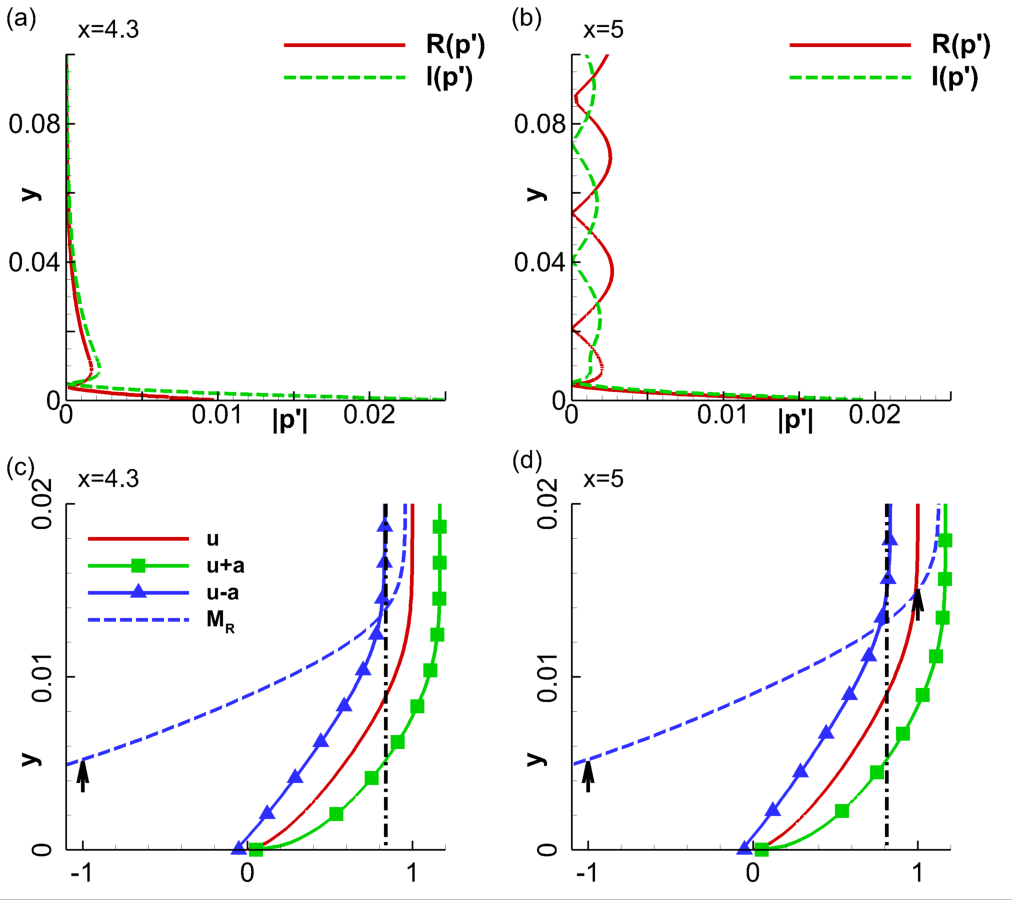}}
\caption{Absolute values of real and imaginary components of the pressure eigenfunction, $R(p')$, and $I(p')$, respectively, plotted at (a) $x=4.3$ - prior to synchronization and (b) $x=5$ - after synchronization. 
Profiles of various speed limits as indicated, plotted at (c) $x=4.3$ and (d) $x=5$. 
The vertical dashed-dotted lines in (c) and (d) mark the phase speed of the corresponding unstable wave.}
\label{figegmen}
\end{figure}

The behavior of the unstable eigenmode is qualitatively different prior to and after the synchronization of mode~F phase speed with $1-1/M_\infty$.
This is demonstrated in figure~\ref{figegmen}, which plots  the pressure eigenfunctions at two locations, $x=4.3$ (before synchronization) and $x=5$ (after synchronization), for $T_W=0.1$ in  figures~\ref{figegmen}(a) and (b), respectively.  
Prior to the synchronization, the eigenfunction has the typical compact form with a single zero-crossing associated with the second-mode instability \citep{mack1984boundary}. 
Once the phase speed of the instability wave falls below $(1-1/M_\infty)$, the outer region exhibits an oscillatory behavior. 
The phase speed relations at these two locations are further examined in figures~\ref{figegmen}(c) and (d). 
Profiles of mean velocity ($u$), fast acoustic ($u+a$) and slow acoustic ($u-a$) speeds, and the relative Mach number ($M_R=(u-c)/a$) are plotted together with the phase speed of the instability wave, $c_r$, represented by a vertical dashed-dotted line. 
The profiles show that  at $x=4.3$, $M_R < 1$ at all wall-normal locations.
Since the wave is then subsonic relative to the mean flow in the outer boundary layer, it decays into the freestream. 
However, it is supersonic relative to the mean flow near the wall ($M_R < -1$), in the region, $0 \le y \le 0.0052$; this location is marked by the vertical arrow which indicates the sonic limit, $M_R = -1$. 
At $x=5$, the lower supersonic region remains similar to that observed prior to synchronization. 
The eigenfunctions below this sonic limit also appear similar in figures~\ref{figegmen}(a) and (b). 
However, $M_R$ now exceeds the sonic limit in the outer boundary layer as well, which is also marked by a second vertical arrow at $y \sim 0.015$ (figure~\ref{figegmen}(d)), where $M_R = 1$. 
This relative supersonic region results in wave radiation into the freestream, consistent with the oscillatory eigenfunction. 

\subsection{Physical nature of eigenmodes}\label{sec_paem}
The above eigenspectrum analysis provides a mathematical description of the behavior of instabilities. 
In such scenarios, where the qualitative behavior of instabilities varies in a meaningful fashion, a physics-based interpretation can provide complementary insights into key flow mechanisms \citep{tumin2020lst}. 
For this, we adopt a Kov{\'a}sznay-type framework, which decomposes fluctuations in terms of three physical components: vortical, acoustic and entropic, referred to earlier as fluid-thermodynamic (FT) components. 
This framework is used to study the composition of instabilities in the vicinity of synchronization; this facilitates a better understanding of the physical changes induced in mode~F due to wall cooling..

The momentum potential theory (MPT) of \citet{doak1989momentum} provides an effective way to achieve the desired splitting without linearization.
Specifically, it can be shown that the ``momentum-density" vector, $\rho\boldsymbol{u}$, can be split precisely into rotational and irrotational mean and fluctuating components.
%, regardless of the magnitude (or nonlinearity) of the fluctuation.
%because it appears naturally in linear form in the continuity equation.
The former represents the vortical component while the latter can be (again exactly) further decomposed into acoustic (irrotational and isentropic) and entropic (irrotational and isobaric) components. 
Details of the procedure, and implementation for LST as well as DNS data, may be found in \citet{unnikrishnan2016acoustic,unnikrishnan2018transfer} for free shear layers and \citet{unnikrishnan2019interactions} for wall bounded layers.
Briefly, the decomposition can be expressed as follows:
\begin{equation}\label{momsplit}
\rho \boldsymbol{u} = \overline{\boldsymbol{B}} + \boldsymbol{B}' - \nabla \psi', 
\qquad \nabla\boldsymbol{.}\overline{\boldsymbol{B}}=0, \qquad \nabla\boldsymbol{.B}'=0.
\end{equation} 
where $\overline{\boldsymbol{B}}$ and $\boldsymbol{B}'$ are the mean and fluctuating parts of the vortical component, respectively.
The irrotational component,$-\nabla \psi'$ may be further split to obtain the acoustic ($-\nabla \psi'_A$) and vortical ($-\nabla \psi'_T$) components through the following Poisson equations \citep{unnikrishnan2019interactions}:
\begin{equation}\label{poitot}
\nabla^2 \psi'=\frac{\partial \rho'}{\partial t}, 
\qquad
\nabla^2 \psi'_A = \frac{1}{a^2}\frac{\partial p'}{\partial t}.
\end{equation}
\begin{equation}\label{pois_psiat}
\psi'=\psi'_A + \psi'_T.
\end{equation}
Here $t$ is time, and $a$ is the local sonic speed. 
For convenience, we adopt the following notation for the FT variables in 2D space: $\boldsymbol{B}'=(B_x',B_y')$,  $-\nabla \psi_{A}'=\boldsymbol{A}'=(A_x',A_y')$, and $-\nabla \psi_{T}'=\boldsymbol{\tau}'=(\tau_x',\tau_y')$. 

\begin{figure}
\centering
\setlength\fboxsep{0pt}
\setlength\fboxrule{0pt}
\fbox{\includegraphics[width=5.0in]{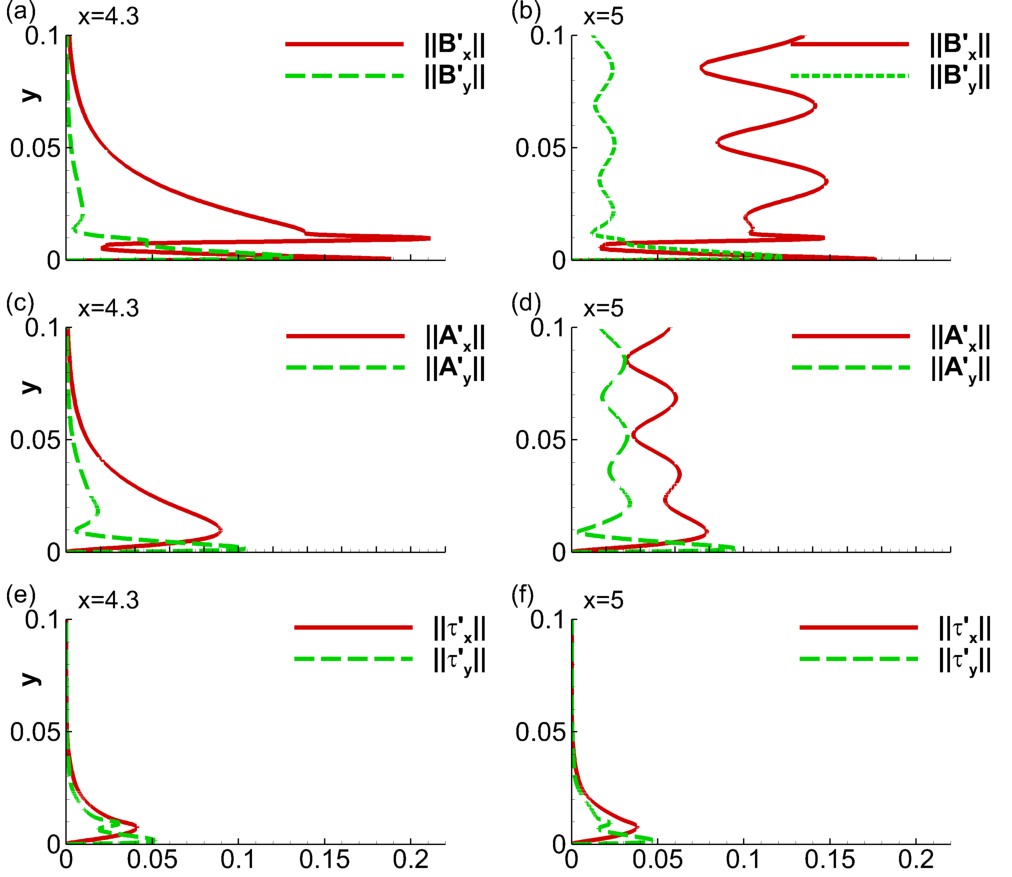}}
\caption{Modulus of streamwise and wall-normal components of (a) vortical, (c) acoustic and (e) entropic constituents of the unstable eigenmode at $x=4.3$. 
(b), (d) and (f) - corresponding results at $x=5$.}
\label{figfteig}
\end{figure}

The decomposition of cold wall eigenfunctions shows very different behavior from those of the adiabatic wall previously described in \citet{unnikrishnan2019interactions}. 
The extracted FT components of the eigenfunctions of the coldest wall, $T_W = 0.1$, are analyzed  prior to and after synchronization in figure~\ref{figfteig}, using the modulus of the streamwise and wall-normal components.
The streamwise locations correspond to those in figure~\ref{figegmen}.
%The extracted FT components are plotted in figure~\ref{figfteig}, 
The top, middle and bottom rows correspond to the vortical, acoustic and entropic components, respectively. 
The left and right columns display the FT decomposition of the eigenfunctions before and after the synchronization. 
The FT composition prior to synchronization indicates that the vortical component is the dominant one in the Mack mode instability over highly-cooled walls, followed by the acoustic and entropic components. 
This hierarchy is consistent with corresponding results for an adiabatic wall \citep{unnikrishnan2019interactions}. 
Post-synchronization however, the oscillatory eigenfunction exhibits an acoustic component in the freestream, as seen in figure~\ref{figfteig}(d) -- this represents the observed acoustic radiation.
Somewhat surprisingly, and confirmed later with DNS results, the FT decomposition also reveals that the radiated waves in the freestream contain a significant amount of vorticity, as seen in figure~\ref{figfteig}(b). 
Streamwise vortical fluctuations are considerably higher than the wall-normal component, and this difference amplifies in the downstream direction, as the wave-fronts become increasingly inclined towards the wall. 
The contribution from the entropic component to freestream radiation remains negligible. 
Thus, the emission from cold HBLs resulting from supersonic phase speeds of instabilities contains comparable amounts of vortical and acoustic waves that propagate in the freestream. 
Having characterized the linear mechanisms, we next quantify the perturbation growth and energy efflux in these HBLs, by adopting a high-fidelity numerical approach.

\section{Analyses of linear perturbations with 2D DNS}\label{sec_lnpran}
The above results show that wall cooling can introduce significant variations in perturbation growth characteristics of HBLs. 
In this section, the relative behavior of the different cold-wall HBLs is examined for all cases listed in  table \ref{tab:casdcr},  by adopting  DNSs  to capture the spatio-temporal evolution of second-mode instabilities.

\subsection{Laminar and forced DNSs}\label{sec_lmfrdns}
The simulations solve the 2D, unsteady Navier-Stokes equations in curvilinear coordinates.  
To ensure sufficient resolution, a seventh-order WENO \citep{balsara2000monotonicity} scheme is employed to reconstruct the characteristic variables.
The inviscid fluxes are then obtained using the Roe scheme \citep{rpl81-1}, along with an entropy fix \citep{egorov2006direct}.  
Fourth-order central differences are used to discretize the viscous terms.  
Time integration is performed using a nonlinearly stable third-order Runge-Kutta scheme \citep{shu1988efficient}.  
To improve the robustness of the solver, the high-order reconstruction is substituted by a third-order upwind scheme in the vicinity of shocks \citep{bhagatwala2009modified}, along with the van-Leer harmonic limiter \citep{lbv79-1}, to minimize grid-scale oscillations.

The computational domain spans $0\leq x \leq 7$ and $0\leq y \leq 1.32$, thus including the leading edge of the plate ($x=0$) in the calculations. 
Supersonic inflow and outflow conditions are imposed on the upstream and downstream boundaries, respectively.
Zero-normal-gradients are imposed on the freestream boundary. 
Wall cooling is simulated using isothermal conditions on the surface. 
%Following the convergence of the mean, perturbations are introduced in the form of harmonic wall-normal momentum perturbations to emulate a wall blowing-suction actuator \citep{egorov2006direct,soudakov2010numerical}. 
For each case, the basic state is first obtained and subsequently, perturbations are introduced in the form of harmonic wall-normal momentum perturbations to emulate a wall blowing-suction actuator \citep{egorov2006direct,soudakov2010numerical}. 
The actuation in this set of forced DNSs has the form:
\begin{equation}
q_w(x,t)= \rho_w v_w =A_{2D}sin\left(2\pi\frac{x-x_1}{x_2-x_1}\right)sin(\omega_{2D} t).
\end{equation}\label{eqn_wbs2d}
Here, $q_w$ is the wall-normal momentum, $A_{2D}$ is the forcing amplitude, and $x_1$ and $x_2$ define the streamwise extent of the actuator, $x_1 \leq x \leq x_2$. 
For all wall temperatures, the actuator is placed at,  $0.5\leq x \leq 0.525$, to ensure that it is upstream of the region of second-mode instability.
The linear stability analyses of \S~\ref{sec_lsrls} reveal that in the chosen streamwise extent of the computational domain,  a wave with circular frequency, $\omega \sim 150$, exhibits linear amplification due to second-mode instability in all cases. 
Hence, this frequency is chosen to perturb all the mean flows ($\omega_{2D}=150$). 
The forcing amplitude is chosen as $A_{2D}=1\times10^{-4}$, to introduce perturbations in the linear range. 

\begin{figure}
\centering
\setlength\fboxsep{0pt}
\setlength\fboxrule{0pt}
\fbox{\includegraphics[width=5.0in]{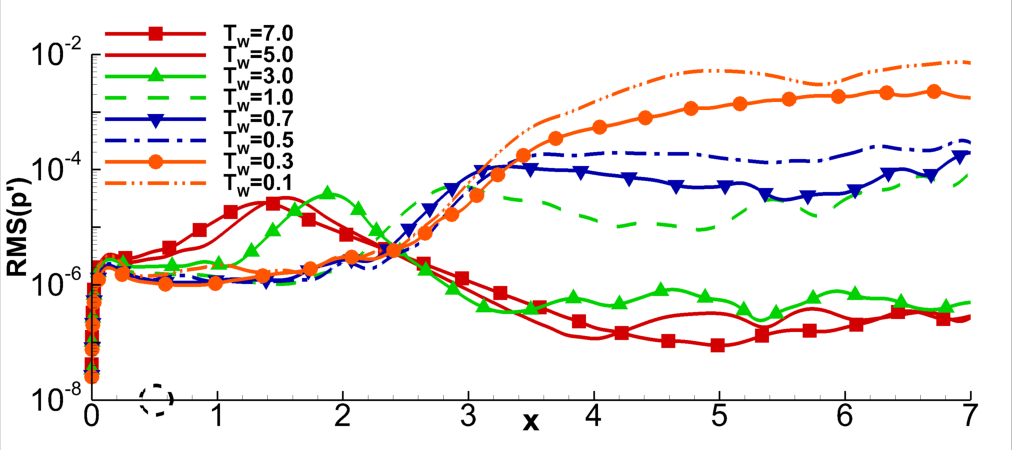}}
\caption{Streamwise variation of RMS values of wall-pressure perturbations for various cold walls. 
The dashed circle on the horizontal axis marks the streamwise location of the actuator.}
\label{figprmsr1}
\end{figure}

Cooling destabilizes the second-mode, and the resulting higher growth rates increase the susceptibility of the HBLs to transition \citep{stetson1989laminar,stetson1992hypersonic}. 
Wall-pressure perturbations provide a means to quantify this trend in the HBLs examined.
The streamwise variation of wall-pressure perturbations is presented in terms of root-mean-square (RMS) values in figure~\ref{figprmsr1} for all eight cases. 
The streamwise location of the actuator is also marked on the horizontal axis with a dashed circle. 
As expected, the warmer-walls experience relatively weak amplification of perturbations. 
The peak values achieved increase with wall cooling, consistent with the destabilization of Mack mode. 
The region of amplification also progressively shifts downstream, due to the thinning of the boundary layer in cooler HBLs. 
The peak amplitudes vary by over three orders of magnitude across the cases.
For the warmer-walls, the RMS perturbations attenuate outside the upper branch of the neutral curve. 
For highly cooled walls (typically $T_W \le 1$), this attenuation is weak, and the high amplitudes are sustained over a prolonged region downstream. 
This could be the result of the merging of the weakly-damped discrete mode with the continuous slow acoustic spectrum, resulting in inter-modal interactions, and is consistent with the observations in \citet{chuvakhov2016spontaneous}.

\begin{figure}
\centering
\setlength\fboxsep{0pt}
\setlength\fboxrule{0pt}
\fbox{\includegraphics[width=5.0in]{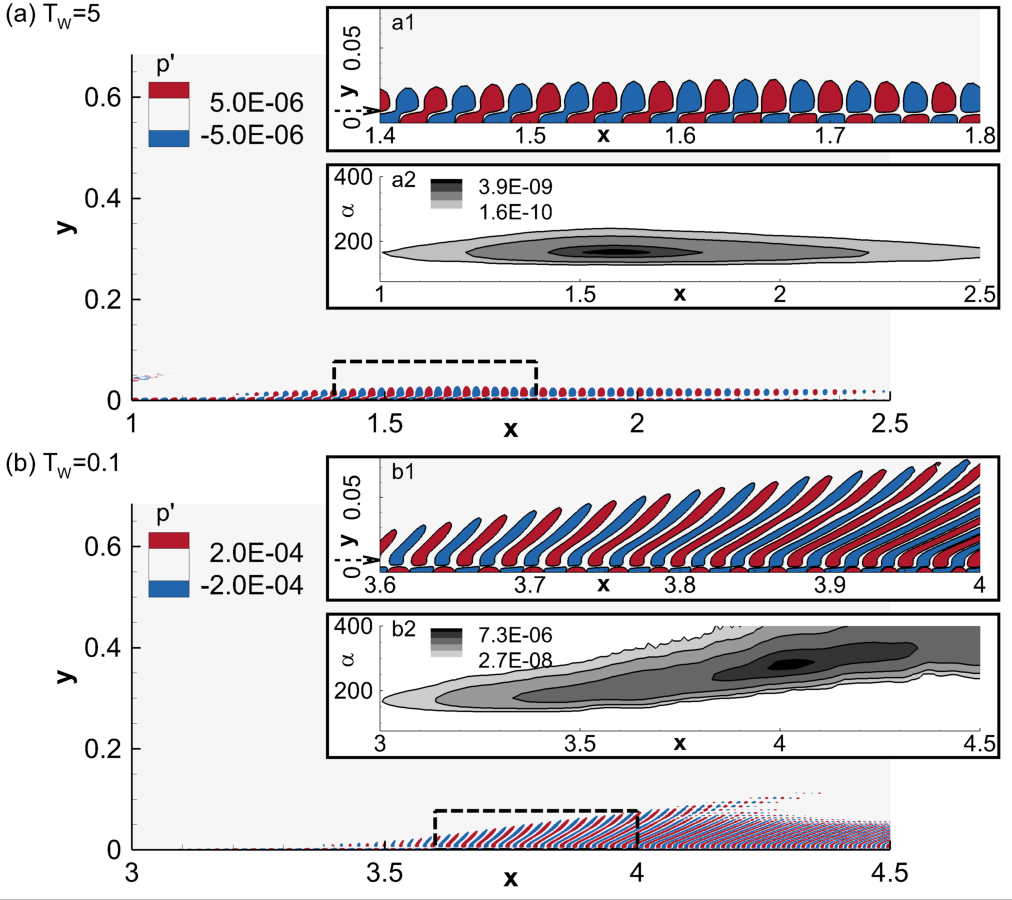}}
\caption{Pressure perturbation contours highlighting the form of second-mode instability for (a) $T_W=5$ and (b) $T_W=0.1$. 
The regions marked by the dashed rectangles in (a) and (b) are magnified in the respective insets, a1 and b1. 
Insets a2 and b2 plot the space-wavenumber, $x-\alpha$, variation in the pressure perturbation signal, for $T_W=5$ and $T_W=0.1$, respectively, at a wall-normal location, $y \sim 0.008$. 
For reference, this location is marked with horizontal dashed arrows near the y-axes of insets a1 and b1.}
\label{figppct}
\end{figure}

Key features of pressure perturbations in the second-mode instability in these HBLs are summarized in figure~\ref{figppct} using the pressure perturbation fields for $T_W=5$ (figure~\ref{figppct}(a)) and $T_W=0.1$ (figure~\ref{figppct}(b)). 
For each case, the region of second-mode amplification is displayed, and magnified views of the wave-forms are provided in the respective insets (a1 and b1). 
The warmer plate shows the classic second-mode lobes with a zero-crossing near the GIP, and a compact wall-normal profile. 
For the cold wall however, these waves extend into the freestream at an acute angle relative to the plate.
This angle decreases in the downstream direction, causing the wave-fronts to align increasingly towards the wall. 

This behaviour is consistent with the reduction of the phase speed of the instability, which is now more supersonic with respect to the freestream; as such, the waves propagate further out into the farfield.
Since the mechanism is linear, an accompanying continual increase in the wavenumber of the instability wave is observed. 
This is quantified with scalograms provided in the second inset of each panel of figure~\ref{figppct} (a2 and b2), which track the variation in streamwise wavenumber, $\alpha = 2 \pi / \lambda$, with $x$, at a wall-normal location, $y \sim 0.008$,  marked for reference by a horizontal dotted arrow on the vertical axes in insets a1 and b1.
Over the warmer-wall, the instability wave maintains a near-constant wavenumber throughout the unstable region. 
The situation is clearly very different over the cooler-wall, where the wavenumber increases as the wave amplifies in the HBL. 

\subsection{Radiating modes and efflux of energy}\label{sec_rmefle}
The efflux of energy from radiating HBls may have the potential to materially impact the growth rates of perturbations, which in turn may present options to aid transition mitigation \citep{chuvakhov2016spontaneous}. 
FT dynamics is now employed to examine this physics in the region of instability growth captured in the DNSs. The procedure followed is to first identify the physical mechanisms being affected due to wall cooling, and to then utilize this knowledge to quantify the energy radiated from the HBL. 

\begin{figure}
\centering
\setlength\fboxsep{0pt}
\setlength\fboxrule{0pt}
\fbox{\includegraphics[width=5.0in]{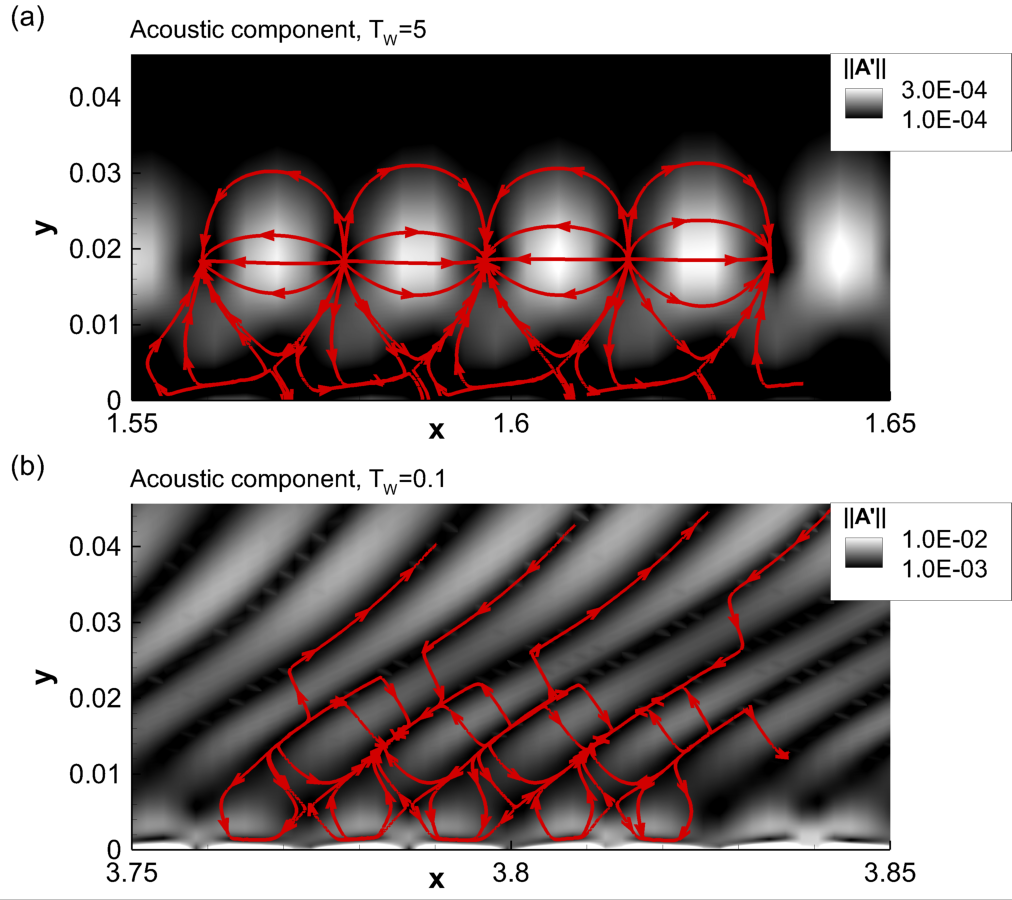}}
\caption{Flux lines of the acoustic component, $(A_x',A_y')$, for (a) $T_W=5$ and (b) $T_W=0.1$. 
The contours represent the magnitude of the acoustic field.}
\label{figaccont}
\end{figure}

Mack mode instabilities relevant to this study are essentially inviscid instabilities, that are generally accepted to exhibit a waveguide (or trapped) behavior \citep{fedorov2011transition} under adiabatic wall conditions, imparting them an acoustic character.
A clearer understanding of how the properties of the trapped component changes when the wall is cooled, and the boundary layer starts radiating, aids the subsequent energy efflux study. 
Prior FT analysis of Mack modes in HBLs with adiabatic walls ($T_w \sim 7$) \citep{unnikrishnan2019interactions} has revealed that both mode~F and mode~S instabilities are composed of all three FT components (vortical, acoustic and entropic). 
However, the above waveguide behavior is most clearly manifested in the acoustic component \textit{i.e.,} flux lines associated with $\boldsymbol{A}'$ display alternating monopoles along an essentially horizontal line at approximately the height of the critical layer (where $u=c$).
The warmer-walls show the same behavior; an example is shown by plotting aspects of the acoustic component for $T_W=5$ in figure~\ref{figaccont}(a).
The streamwise extent is chosen to represent a detailed view of the corresponding cases described previously in figure~\ref{figppct}. 
The background contours represent the magnitude of the acoustic component, $\|{\boldsymbol{A}'}\|$. 
The arrow-heads mark the direction of the flux lines which constitute the acoustic component of momentum fluctuations, $(A_x',A_y')$, in the Mack mode. 
Note that since the decomposed variable is $\rho \boldsymbol{u}$, these effectively represent mass fluxes associated with the acoustic component.
For the moderately cooled wall, $T_W=5$, the acoustic flux lines are generally similar to those in the adiabatic case, with a zero-flux line existing at $y \sim 0.02$, corresponding to the critical layer of this instability wave. 
Clearly, this component is effectively trapped between the wall and the critical layer, since no flux lines cross the line joining the monopoles.
%For this, forms of the acoustic component in Mack modes (extracted using MPT) at two levels of wall-cooling are provided in figure~\ref{figaccont}, 

The same results for the highly cooled wall, $T_W=0.1$, are  shown in figure~\ref{figaccont}(b).
Clearly, no zero-flux line exists here, and the arrows indicate that the supersonic mode displays flux transport between the HBL and the freestream due to the acoustic component. 
This is a manifestation of the acoustic part of the radiation associated with the supersonic mode.
Since the vortical ($\boldsymbol{B}'$) and acoustic ($\boldsymbol{A}'$) fluctuations educed from DNSs are inherently components of the momentum vector,  they can transport energy in the flowfield.
These energy considerations, and a budget equation for transport of total fluctuating enthalpy by the different FT components, have been extensively detailed for the adiabatic HBL in \citet{unnikrishnan2019interactions} and for high-speed jets in \citet{unnikrishnan2016acoustic}. 
%Specifically, a budget equation can be written to when specific source terms are activated; a detailed discussion of these energy fluxes transported by the FT components, and the corresponding source terms can be found in \citet{doak1998fluctuating}, along with the governing budget equation. 
%Its application to HBLs have been presented in 

A quantification of the efflux of fluctuating energy associated with each FT component from the HBL requires a splitting of the TFE into those components. 
Such a splitting has been provided by \citet{jenvey1989sound}. 
%, which was employed for adiabatic HBLs in \citet{unnikrishnan2019interactions}.
Since the cold walls considered in this work display radiating modes containing vorticity and acoustic waves in the freestream (see figures~\ref{figfteig}(b) and (d)), both fluxes of the respective energies are of principal importance here.
Specifically, the transported energy variables that must be accounted for are $H_B'$ and $H_A'$, which are the vortical and acoustic components, respectively, of $H'$ and are defined as:
%, which  is the fluctuating component of the enthalpy per unit mass, $H$. 
%Similarly, $H_A'$ is the acoustic component of TFE. 
\begin{equation}\label{eqa}
\begin{split}
H& = c_pT +{\boldsymbol{u.u}}/{2}, \quad H=\overline{H} + H', \\[3pt]
H_B'& = \left(\frac{\overline{c}}{\overline{\rho}}\right)\overline{\boldsymbol{M}}\boldsymbol{.B}', \\[3pt]
H_A'& = \left(\frac{p'}{\overline{\rho}}\right)\left(1-\overline{\boldsymbol{M}}\boldsymbol{.}\overline{\boldsymbol{M}}\right) - 
\left(\frac{\overline{c}}{\overline{\rho}}\right) \overline{\boldsymbol{M.}}\nabla\psi'_A,
\end{split}
\end{equation}
%In the above, $c_p$ is the specific heat at constant pressure, $\boldsymbol{u}$ is the Cartesian velocity vector, and $\boldsymbol{M}\equiv{\boldsymbol{u}}/{c}$ is the Mach number vector. 
where $\boldsymbol{M}\equiv{\boldsymbol{u}}/{c}$ is the Mach number vector. 
The vortical flux of TFE in the wall-normal direction is calculated as $F_B = H_B'\boldsymbol{B}' \cdot \hat{j}$ and, similarly, the acoustic flux is $F_A = H_A'\boldsymbol{A}' \cdot \hat{j}$, where  
$\hat{j}$ is the unit normal in the wall-normal direction. 

%  Integrate

%\textbf{DVG Need to integrate: In the present work, we evaluate only the flux terms in this budget equation, to quantify the efflux of fluctuating energy from the HBLs. }

%% end integrate
Accurate quantitative comparisons of perturbation energy fluxes require care in establishing consistency across all eight boundary layers examined.
Indeed, use of the DNSs described above for this purpose may be misleading, since the actuators are placed at the same streamwise location, but the neutral curves depend on the wall temperature. 
To ensure that the linear amplification is initiated at similar perturbation energy levels in each case, a second set of forced DNS is performed for all cases, with actuators placed at similar locations relative to the lower branch of the respective neutral curve. 
These locations can be identified for each mean flow using growth rates calculated from the amplitude of wall-pressure perturbations obtained from the first set of forced DNSs discussed above. 
To minimize numerical noise contamination, the neutral curve is assumed to begin at the threshold, $\sigma = {d \{ln [a_p(x)]\}}/{dx} \sim 2$ \citep{egorov2006direct}, where 
$\sigma$ is the numerical growth rate and $a_p(x)$ is the wall-pressure amplitude. 
Thus, for this second set of forced DNSs, the streamwise extent of the actuator is defined as $x_1=x_0$ and $x_2=x_1 + 0.025$, where $x_0$ is the lower neutral limit of the respective HBLs. 
All other actuator parameters in (\ref{eqn_wbs2d}) are identical to the first set of forced DNSs. 
To prevent superfluous computations, the domain upstream of the neutral point is neglected by imposing corresponding laminar profiles at the inlet boundary.
%, in the second set of forced DNSs. 
The outflow boundaries are placed approximately $4$ nondimensional length units downstream to ensure that the linear instability range is accommodated. 
The perturbation evolution and energy fluxes are analyzed over this streamwise distance.

\begin{figure}
\centering
\setlength\fboxsep{0pt}
\setlength\fboxrule{0pt}
\fbox{\includegraphics[width=5.0in]{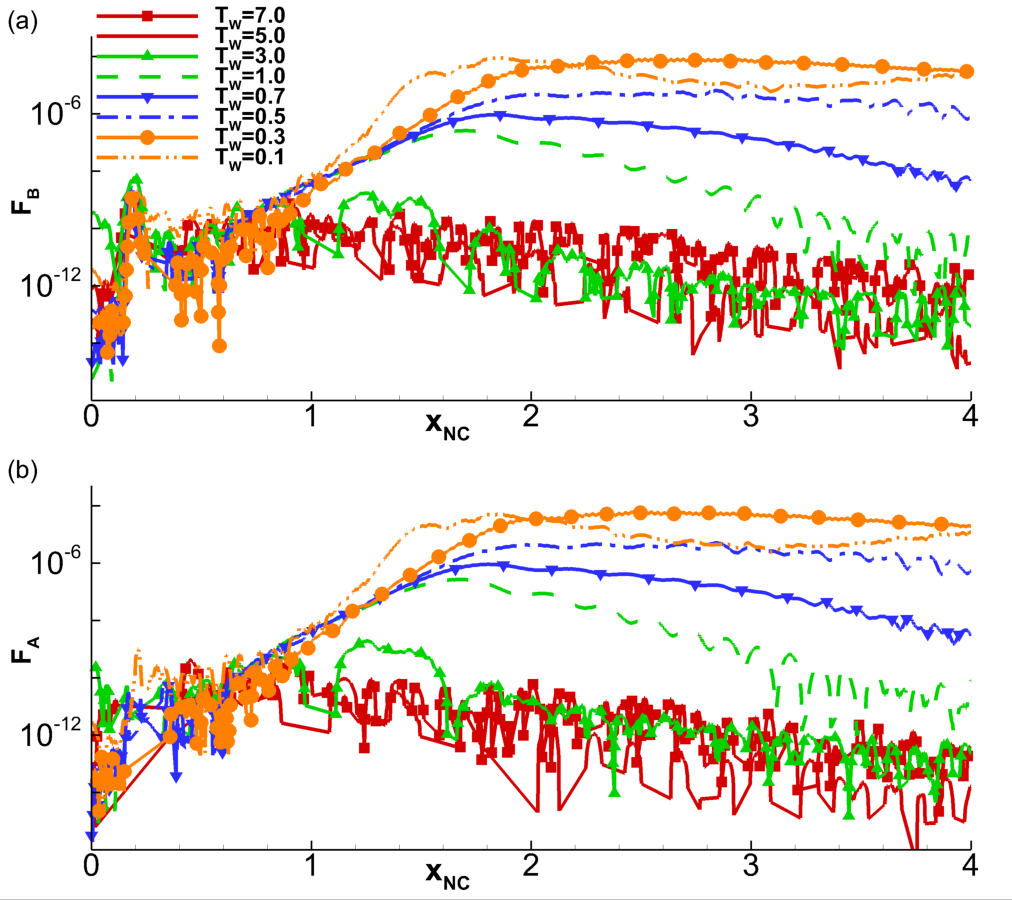}}
\caption{Streamwise variation of (a) vortical and (b) acoustic efflux of energy from the HBLs, for various cold walls. 
The actuator is placed at the lower neutral limit.}
\label{fignucrflx}
\end{figure}

The hydrodynamic and acoustic energy fluxes, $F_B$ and $F_A$ thus obtained are plotted in figure~\ref{fignucrflx}.  
Note that the streamwise direction is now represented using the displaced coordinate, $x_{NC}=x-(x_0 + \Delta x)$, with origin near the lower neutral limit. 
$\Delta x=0.2$ is a small upstream displacement of the inlet boundary from this neutral limit, which is included to prevent the actuator from interfering with the inflow boundary conditions. 
For all cases, the fluxes are calculated at a surface located at $y \sim 0.04$, which is sufficiently outside the boundary layer. 
The positive values of fluxes indicate that $F_A$ and $F_B$ both constitute a net outflow of energy from each HBL.
More interestingly, both these fluxes are of comparable magnitudes for a given wall temperature. 
This reinforces the observation made in the context of LST above, that vortical and acoustic mechanisms are equally significant in radiation processes from cooled HBLs. 
The radiative loss of energy is negligible for near-adiabatic and moderately cooled walls. 
The energy loss increases monotonically with wall cooling; in fact, when $T_W \le 1$, the energy efflux is between $5$ to $6$ orders of magnitude higher than for moderately cooled wall boundary layers. 
The radiation from these highly-cooled wall layers exhibits peak values slightly downstream of the location where the amplitudes of the linear waves achieve their maximum values on the surface. 
This is due to the forward inclination of the supersonic waves in the freestream, as observed earlier in figure~\ref{figppct}(b). 

\begin{figure}
\centering
\setlength\fboxsep{0pt}
\setlength\fboxrule{0pt}
\fbox{\includegraphics[width=5.0in]{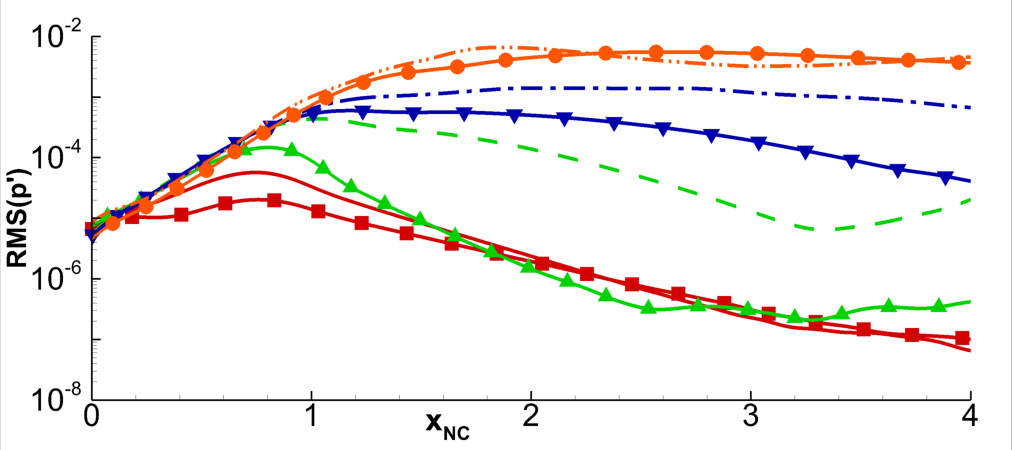}}
\caption{Streamwise variation of RMS values of wall-pressure perturbations for various cold walls, with actuator placed at the lower neutral limit.}
\label{figprmsnucr}
\end{figure}

The efflux of vortical and acoustic enthalpy diverts a portion of fluctuating energy from the primary second-mode instability mechanism. 
To evaluate the implication of the efflux on transition, the RMS of wall-pressure perturbations is shown in figure~\ref{figprmsnucr} for this second set of forced DNSs.
A main observation emerging from the figure is that the increases in radiative losses do not have a prominent impact on wall-pressure perturbations, whose value at any given distance increases monotonically with wall-cooling. 
This indicates that the efflux due to the supersonic radiating mode is a higher-order effect and may not be sufficient to compensate for the first-order destabilizing effect of wall cooling on the second-mode.
This observation is not inconsistent with recent theoretical analysis by \citet{tumin2020wave}, who  notes that the radiating component is limited to the ``tail'' portion of second-mode wavepackets in cooled HBLs, and do not attenuate the primary amplification mechanism associated with the subsonic component.

\section{3D transition mechanisms in cooled HBLs}\label{sec_trnchbl}
The above analysis addressed the fundamental aspects of stability modes and linear growth through LST and 2D DNSs.
The problem of breakdown associated with transition requires a 3D analysis; such an exercise is now conducted with fully 3D DNSs to describe the effects of radiating modes on transition in wall-cooled  HBLs. 
%transition, in the presence of radiating modes.
Specifically, the evolution of spatio-temporally localized wavepackets is examined for representative cases as noted in table \ref{tab:casdcr}. 
%Note that 3D-DNS results are performed for most, but not all, cases to limit computational expense. 
%This is enabled through 3D high-fidelity simulations of these HBLs, subjected to relevant forcing conditions. 
Below, the numerical framework is first summarized, followed by a comprehensive qualitative and quantitative spectral and statistical description of the associated dynamics, including effects on wall loading.

\subsection{General features of 3D response to wavepacket forcing}\label{sec_3dwpp}
The 3D Navier-Stokes equations are solved using the high-order spatial schemes summarized in \S\ref{sec_lmfrdns}. 
To facilitate feasible time-step-sizes, the second order diagonalized \citep{pth81-1} implicit Beam-Warming scheme is utilized for time-integration \citep{br78-1}. 
The extent of the computational domain for each HBL in the 3D DNS is as follows: 
\begin{equation}
x_0 - \Delta x \leq  x \leq x_0 - \Delta x + 4.1, \quad
0 \leq y \leq 0.43, \quad
-0.15 \leq z \leq 0.15.
\end{equation}
As discussed in the context of the second set of forced 2D DNS employed for energy efflux considerations, $x_0$ is the lower neutral limit of the respective HBL, and $\Delta x = 0.2$. 
The computational domain is composed of $4001$, $261$ and $151$ nodes in the streamwise, wall-normal and spanwise directions, respectively, for all the wall temperatures reported here. 
Around $45\%$ of the points are clustered within the boundary layer height of the coldest wall, $T_W=0.1$. 
A relatively finer wall spacing of $\Delta y_W = 8 \times 10^{-5}$ is utilized to ensure sufficient resolution of the cooler boundary layers, based on prior considerations for near-adiabatic wall temperatures \citep{unnikrishnan2019first}. 
The boundary conditions are similar to those used in the 2D simulations at the inflow, outflow, wall and farfield boundaries, and the spanwise direction is assumed to be periodic. 
3D wavepackets are introduced though wall-blowing-suction, using the following actuator model \citep{novikov2016direct,novikov2017direct}:
\begin{equation}
q_w(x,z,t)=\rho_w v_w =
A_{3D}~sin\left(2\pi\frac{x-x_1}{x_2-x_1}\right)
sin\left(\pi\frac{z-z_1}{z_2-z_1}\right)
sin(\omega_{3D} t).
\end{equation}\label{eqn_3dwbsc}
The forcing amplitude of the 3D wavepacket, $A_{3D}=5\times10^{-2}$, and the streamwise extent of the actuator is set to $x_1=x_0$ and $x_2=x_1 + 0.025$. 
The circular frequency of excitation is $\omega_{3D} =150$, and the actuator width in the spanwise direction is bounded by $z_1=-0.01$ and $z_2=0.01$. 
A random background perturbation field with an RMS of $O(1\times10^{-4})$ is also included in the actuator to enhance the stochastic nature of simulations that may reach post-breakdown conditions \citep{sayadi2013direct}. 
To localize the wavepacket in time, the actuator is turned off after $2$ time periods of the wave, by bounding the perturbation-input within the time-interval, $0 \le t \le 4 \pi / \omega_{3D}$. 

\begin{figure}
\centering
\setlength\fboxsep{0pt}
\setlength\fboxrule{0pt}
\fbox{\includegraphics[width=5.0in]{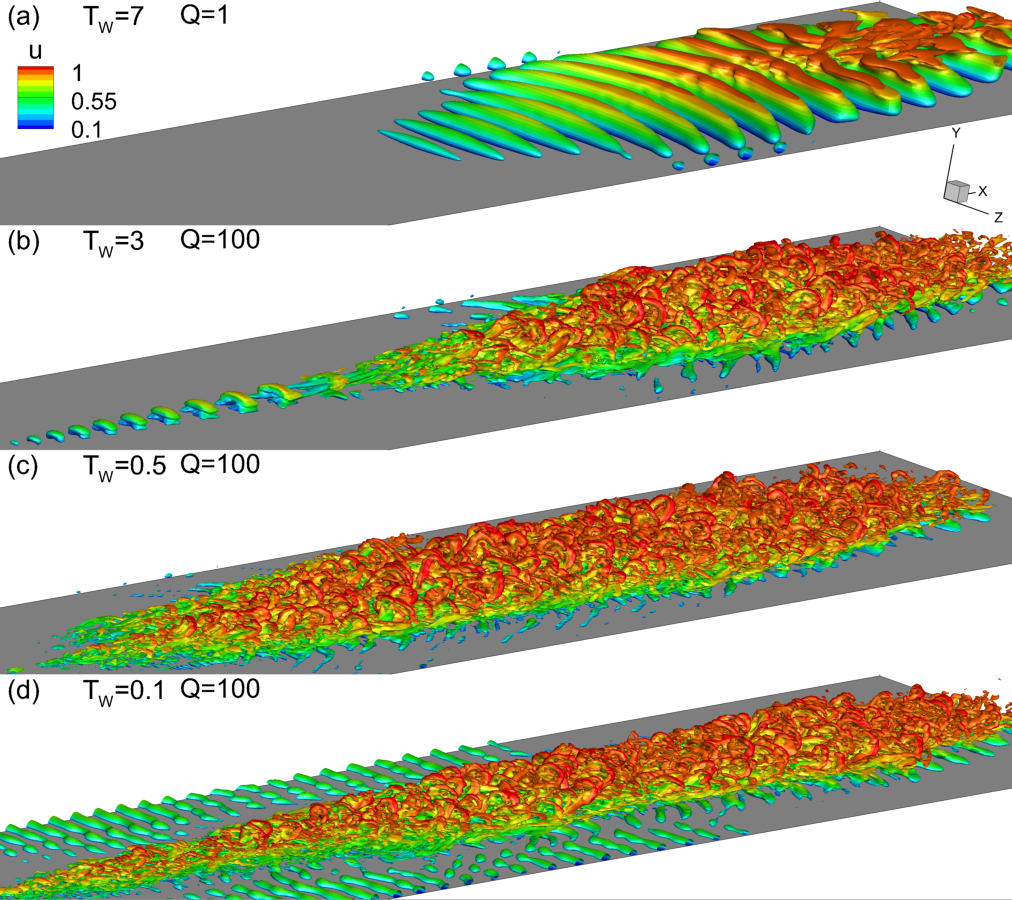}}
\caption{Instantaneous vortical structures induced by the wavepacket, visualized using Q-criterion, colored by $u$. 
Results obtained when the perturbations reach the exit of the computational domain in each case. 
(a) $T_W=7$, (b) $T_W=3$, (c) $T_W=0.5$, and (d) $T_W=0.1$. 
All panels display identical contour levels, as indicated in (a). }
\label{figqiso4case}
\end{figure}

The overall effects of wall-cooling and supersonic modes on transition are first qualitatively summarized in figure~\ref{figqiso4case}.
The eventual state of the wavepacket near the exit of the computational domain ($x_{NC} \sim 4$) is presented for four of the wall temperatures simulated, $T_W=7$, $3$, $0.5$, and $T_W=0.1$.
Vortical structures as defined by the Q-criterion, colored by streamwise velocity, $u$, are used to visualize the wavepacket. 
The wavepacket in the near-adiabatic wall, $T_W=7$, attenuates outside the region of linear instability without tripping the HBL (figure~\ref{figqiso4case}(a)). 
%Indeed, the forcing chosen here purposefully differs from that employed in earlier work \citep{unni???}, where second-mode breakdown is achieved under adiabatic wall conditions, in order to highlight the effect of wall cooling, for which, breakdown is observed as discussed below.
The low value of the Q-criterion ($Q=1$) reflects this state, and the wavepacket retains the signature of 2D waves injected into the flow by the actuator. 
In the moderately cooled wall, $T_W=3$, the second-mode is destabilized due to wall-cooling, but as observed in the context of figure~\ref{fignucrflx}, does not radiate.
This yields stronger amplification of the wavepacket, and it develops into a ``young turbulent'' spot \citep{siva2010numerical} by the end of the domain, as seen in figure~\ref{figqiso4case}(b), with hairpin vortices trailing the leading ``core'' region. 
The vortical structures in this turbulent spot are thus highlighted using a higher threshold, $Q=100$.

The cases $T_W=0.5$ and $T_W=0.1$ account for the impact of radiative efflux of energy. 
Here, the second-mode is even more destabilized, but also radiates vortical and acoustic components of energy into the freestream (as seen in figure~\ref{fignucrflx}). 
Figures~\ref{figqiso4case}(c) and (d) clearly indicate that these highly cooled HBLs undergo transition due to the amplification and breakdown of the respective wavepacket, giving rise to well-defined turbulent spots. 
Despite radiative losses, wall cooling is observed to increasingly enhance the development to turbulence by second-mode wavepackets, resulting in elongated turbulent spots that sustain fine-scale features. 
The turbulent spots also become ``flatter'' over the cooler-walls, since they are confined to thinner boundary layers. 
Although 3D simulations were performed for $T_W=7, 3, 1, 0.5, 0.3$, and $T_W=0.1$, for brevity, key trends are summarized in the following discussion using the most illustrative cases.
For example, many of the inferences regarding transition are reported using only the two extreme cases, $T_W=3$, and $T_W=0.1$, where the respective HBLs trip. 
Additional cases are included selectively to enhance the clarity of those trends. 

\subsection{Waveform evolution and spectra}\label{sec_evspwp}
\begin{figure}
\centering
\setlength\fboxsep{0pt}
\setlength\fboxrule{0pt}
\fbox{\includegraphics[width=5.0in]{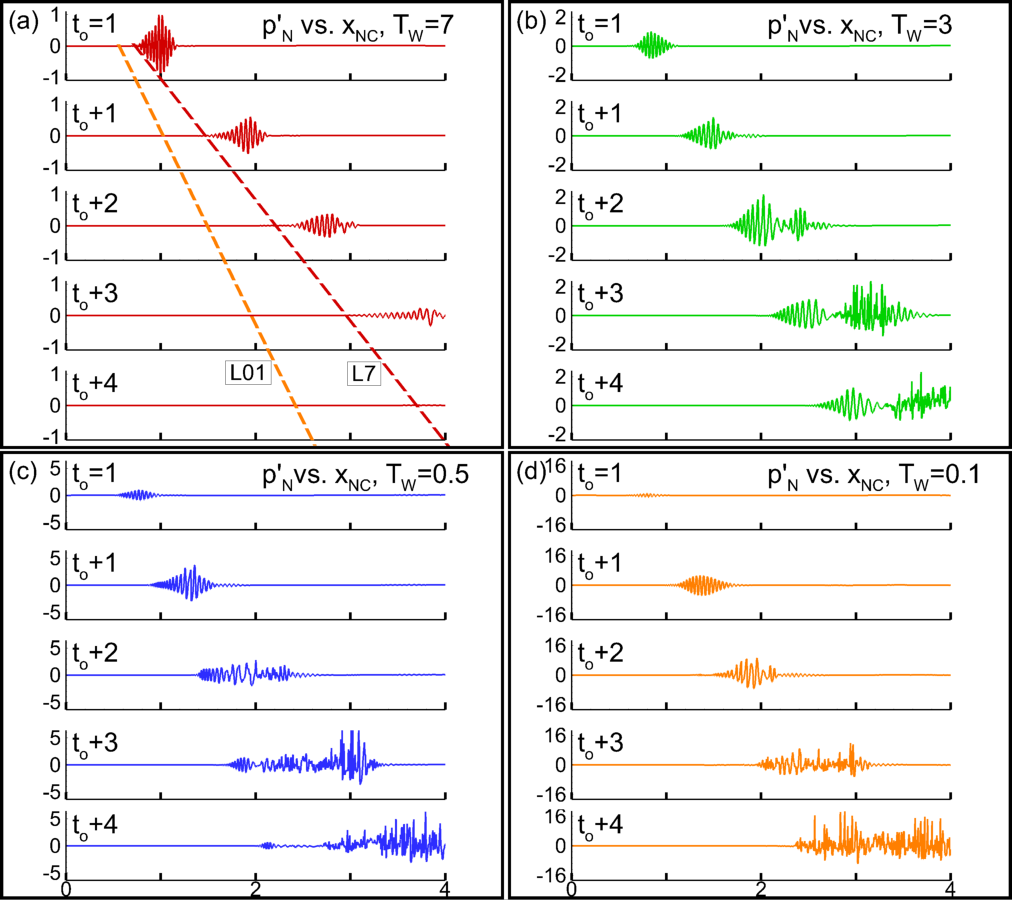}}
\caption{Streamwise distribution of wall-pressure perturbations at the indicated time-instances, for (a) $T_W=7$, (b) $T_W=3$, (c) $T_W=0.5$, and (d) $T_W=0.1$. 
The dashed inclined lines in (a), L01 and L7, trace the trailing edges of the wavepackets in $T_W=7$ and $T_W=0.1$, respectively, to estimate their convective speeds.
}
\label{figpxplt}
\end{figure}

The spatio-temporal evolution of the wavepacket is summarized using wall-pressure perturbations in figure~\ref{figpxplt}, for the cases, $T_W=7, 3, 0.5$, and $T_W=0.1$. 
In each plot, the horizontal axis spans $0 \le x_{NC} \le 4$, and the vertical axis is a relative measure of fluctuating pressure, $p_{N}'$, which is $p'$ normalized by its peak value at the reference time instant, $t_0=1$. 
As observed in the isolevels of figure~\ref{figqiso4case}(a), although the actuator amplitude for $T_W=7$ is large enough to introduce nonlinear perturbations in the HBL \citep{egorov2006direct,ha2019dnsfc}, the forcing profile chosen results in eventual attenuation in the downstream direction, without tripping the HBL. 
$T_W=3$ shows higher amplification in the downstream direction, and by time, $t=t_0+1$, the wavepacket exhibits an early signature of an amplitude modulation at its leading edge. 
In the following snapshots, this leading, modulated envelope develops into a highly nonlinear region, with rapid amplification of perturbations. 
The nonlinear region progressively moves away from the trailing wavepacket, consistent with experimental observations \citep{glezer1989breakdown}, and eventually becomes the ``turbulent core'' \citep{casper2014pressure} of a young turbulent spot. 

Qualitative differences emerge in the evolution of the wavepackets when the wall is highly cooled.
Nonlinear effects modulating the wavepacket-envelopes do not clearly demarcate a leading head and a trailing region. 
This demarcation weakens as the wall is progressively cooled (see figures~\ref{figpxplt}(c) and (d)). 
In these cases, significant streamwise elongation of the wavepackets occurs, and the leading front is followed by several intermittent turbulent regions. 
Cooling also decreases the convective speed of the wavepacket, as is evident by comparing the positions of the leading and trailing edges at different time instances for each wall temperature. 
A rough estimate based on the trailing edge of the wavepacket in the two extreme conditions,  $T_W=7$ and $T_W=0.1$ indicates that the convective speed in the latter case is around $63\%$ that in the former. 
A graphical comparison of these trends is provided in figure~\ref{figpxplt}(a), where $L7$ and $L01$ are the loci of the trailing edges of the wavepacket with $T_W=7$ and $T_W=0.1$ respectively.
A detailed quantification of trends in convective speeds is presented later in \S~\ref{sec_cfwpd}.

\begin{figure}
\centering
\setlength\fboxsep{0pt}
\setlength\fboxrule{0pt}
\fbox{\includegraphics[width=5.0in]{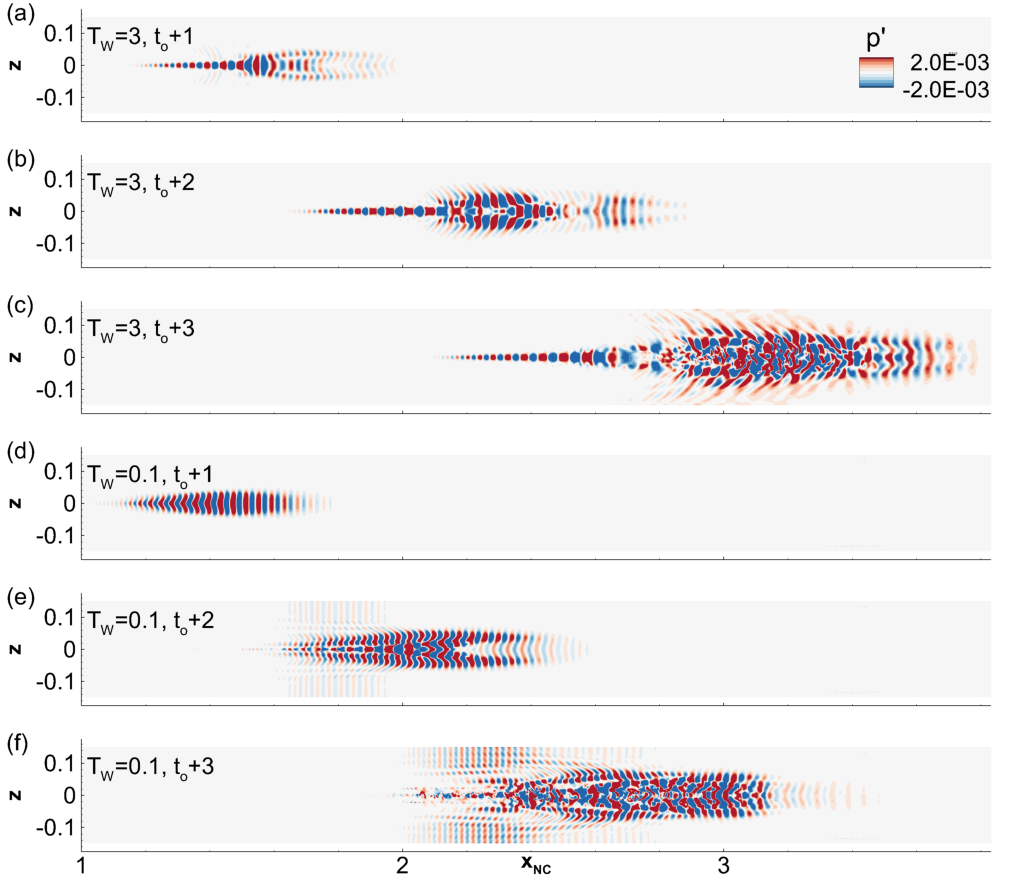}}
\caption{Surface contours of wall-pressure perturbations at the indicated time-instances. 
(a), (b) and (c) correspond to $T_W=3$. 
(d), (e) and (f) correspond to $T_W=0.1$. 
All panels display identical contour levels, as indicated in (a).}
\label{figpxcont}
\end{figure}

The spanwise structure of the wavepackets is now examined using wall-pressure perturbation contours in figure~\ref{figpxcont}. 
To highlight the effect of cooling on wavepacket breakdown processes, two cases, $T_W=3$, and $T_W=0.1$, are chosen at three intermediate time-instances, included earlier in figure~\ref{figpxplt}. 
Comparing figures~\ref{figpxcont}(a) and (d), $T_W=3$ exhibits relatively prominent oblique waves and a nascent head region, whereas $T_W=0.1$ is still dominated by 2D second-mode waves at $t=t_0+1$. 
At the next instant shown (figure~\ref{figpxcont}(b)), the warmer-wall displays a clear demarcation of the head region in the wavepacket, which bifurcates along the center, and contains oblique waves on the outer edges, resulting in lateral spreading. 
This feature was also seen in experimental wall-pressure perturbations of early turbulent spots in a hypersonic nozzle wall by \citet{casper2014pressure}.   
At the corresponding time-instant in the cooler HBL (figure~\ref{figpxcont}(e)), the 2D structures at the center of the wavepacket show early signs of disintegration. 
At $t=t_0+3$, the head region of the warmer wavepacket has already disintegrated into a turbulent spot as seen in figure~\ref{figpxcont}(c), with significant asymmetry about the mid-span. 
Oblique waves also amplify along its edges, widening the spot considerably.

When $T_W=0.1$, although a head region is not conspicuous at $t=t_0+3$, the major portion of the wavepacket undergoes disintegration into an elongated turbulent spot, with little trace of the 2D waves within (figure~\ref{figpxcont}(f)). 
However, unlike for the warmer-wall, this wavepacket contains 2D waves outside its edges, which are prominently observed in $2 \le x_{NC} \le 2.75$. 
This is consistent with predictions of linear theory, according to which 2D waves increasingly amplify with wall-cooling.

\begin{figure}
\centering
\setlength\fboxsep{0pt}
\setlength\fboxrule{0pt}
\fbox{\includegraphics[width=5.0in]{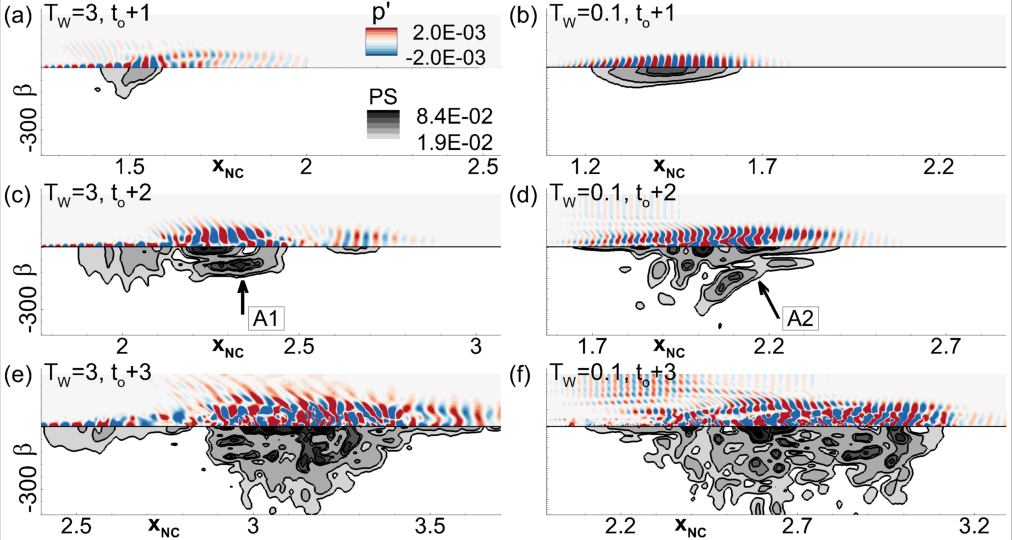}}
\caption{Spanwise wavenumber ($\beta$) spectrum obtained as a function of streamwise distance, at the indicated time-instances. 
(a), (c) and (e) correspond to $T_W=3$. 
(b), (d) and (f) correspond to $T_W=0.1$. 
In each panel, the half-plane of the spectral (vertical) axis is overlaid with the half-plane of the wall-pressure perturbations shown earlier in figure~\ref{figpxcont}. 
Arrows A1 and A2 highlight relevant features in respective panels. 
All panels display identical contour levels, as indicated in (a). }
\label{figpxzwnbcont}
\end{figure}

The corresponding spanwise wavenumber spectra are provided in figure~\ref{figpxzwnbcont}. 
In each panel, only the half-plane of the spanwise wavenumber ($\beta$) axis is included due to its symmetry. 
This is then augmented by a half-plane visualization of the $x-z$ distribution of the wavepacket (discussed above in figure~\ref{figpxcont}), for easy reference. 
For clarity, the instantaneous spanwise spectra are also averaged locally across a time window, $t \pm \Delta t$, where $\Delta t = 0.038$.  
Comparison of figures~\ref{figpxzwnbcont}(a) and (b) confirms a stronger presence of 2D second-mode waves over the cooler plate. 
Following the formation of the head region in the warmer wavepacket, a second energy peak appears at $\beta \sim \pm 93$, as indicated by the arrow A1, in figure~\ref{figpxzwnbcont}(c). 
This is about half the value of the streamwise wavenumber, $\alpha$, of the amplified second-mode instability (see {\em e.g.} inset a2 in figure~\ref{figppct}). 
The cooler wavepacket displays a more continuous range of excited wavenumbers, with peaks at  $\beta \sim \pm 93$ and near its superharmonic, $\beta \sim \pm 196$, which is also marked by arrow A2 in figure~\ref{figpxzwnbcont}(d). 
At later stages of development, both the warmer and cooler wavepackets exhibit a wider range of spanwise wavenumbers (figures~\ref{figpxzwnbcont}(e) and (f), respectively).
However, the cooler wavepacket has relatively more energy content at high wavenumbers, and extends across almost twice the spanwise spectrum of the warmer wavepacket. 
Consequently, the turbulent spots formed over cooler-walls are expected to sustain smaller scales of turbulence, relative to the warmer-wall. 
This trend is further addressed below, in the context of intermittent events in the wake of the head region. 

The spatio-temporal signature of a wavepacket is relatively localized. 
Spectral analyses of such signals are best accomplished by a time-frequency approach, \textit{e.g.}, a scalogram, using a wavelet transformation. 
\citet{yates2020analysis} have recently utilized this technique to identify localized features of stationary cross-flow vortices over a hypersonic cone. 
In addition to its localized nature, perturbations in the wavepacket are also qualitatively different in the leading and trailing regions, which precludes statistical stationarity (in time) of the signal. 
Thus, to quantify the spectral characteristics of wavepacket evolution at different wall temperatures, the scalogram of wall-pressure perturbations is examined at various streamwise locations along the mid-span. 

\begin{figure}
\centering
\setlength\fboxsep{0pt}
\setlength\fboxrule{0pt}
\fbox{\includegraphics[width=5.0in]{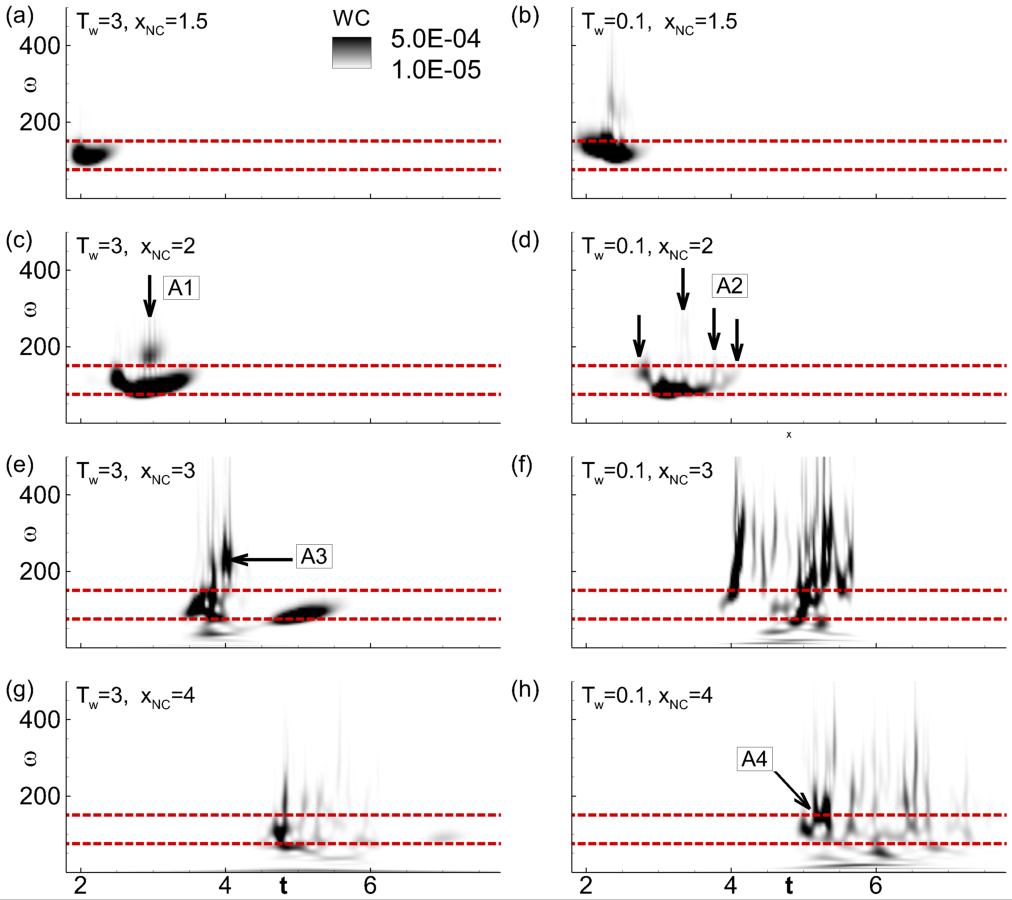}}
\caption{Time-frequency signature of wavepacket evolution represented using scalograms of wall-pressure perturbations on the mid-span, at indicated streamwise locations. 
(a), (c), (e) and (g) correspond to $T_W=3$. 
(b), (d), (f) and (h) correspond to $T_W=0.1$. 
The upper and lower dashed horizontal lines mark the forcing frequency and its subharmonic, $\omega = 150$ and $\omega = 75$, respectively. 
Arrows A1, A2, A3 and A4 highlight relevant features in respective panels. 
All panels display identical contour levels, as indicated in (a). }
\label{figwvltcnt}
\end{figure}

The trends are summarized in figure~\ref{figwvltcnt} using the results for $T_W=3$ (left column), and $T_W=0.1$ (right column). 
The horizontal axis is time, and the vertical axis is circular frequency, \textit{i.e.,}  the pseudo-frequency obtained from the scale-to-frequency conversion relation of the Morlet wavelet used for the transformation. 
Contour values represent the wavelet-coefficient at a given time-frequency coordinate. 
For reference, the forcing frequency ($\omega = 150$) and its subharmonic ($\omega = 75$)  are also marked using dashed lines in each panel. 
The wavepacket is essentially centered around the forcing frequency at upstream locations closer to the actuator. 
Therefore, results are reported at relatively downstream locations, $x_{NC}=1.5, 2, 3$, and $x_{NC}=4$, where nonlinear effects induce significant variations in the spectra. 

At $x_{NC}=1.5$ (figures~\ref{figwvltcnt}(a) and (b))  wavepackets at both wall conditions are still localized, but show the presence of lower frequencies than those of the forcing.
The shift in peak energy is towards the subharmonic, $\omega = 75$. 
Although localized, the cooler-wall HBL already shows early signs of higher frequencies as well, particularly at locations of high wavepacket amplitudes.
Moving downstream, the streamwise elongation in the wavepackets is primarily observed near the subharmonic, as depicted by figures~\ref{figwvltcnt}(c) and (d). 
The warmer-wall wavepacket appears relatively coherent, and the energy at higher frequencies is localized in a zone (marked with arrow A1), which eventually bifurcates into the head region. 
The cooler-wall wavepacket is more dispersive, and higher frequencies are excited at multiple locations, some of which are marked by arrows A2, indicating a stronger presence of nonlinearities. 
At $x_{NC}=3$, figure~\ref{figwvltcnt}(e), the head region of the warmer-wall wavepacket \textit{i.e.,} that appearing earlier on the time axis, is detached from the trailing region, and contains most of the higher broadband frequencies, suggesting its evolution into an early turbulent spot. 
The peak in the higher frequencies at $\omega \sim 230$ (arrow A3) is characteristic of nonlinear interaction between the forcing frequency and its subharmonic, resulting in the quadratic interaction, $\omega_1 + \omega_2 = \omega_3$, with $\omega_1 = 150$, $\omega_2 = 75$, and $\omega_3 = 225$.
The trailing region continues to display a harmonic nature, with a compact frequency distribution. 
The cooler-wall wavepacket does not exhibit a prominent head region (figure~\ref{figwvltcnt}(f)), but contains a series of energetic events, which have broader spectral content than the warmer-wall case. 
This shows that wall cooling results in more rapid development of intermittent turbulence in the wake of the head region, and populates the fine-scale region of the spectrum. 
Further towards the outflow ($x_{NC}=4$), the warmer wavepacket forms a turbulent spot in the leading region, and accounts for most of the energetic perturbations (figure~\ref{figwvltcnt}(g)).  
The cooler wavepacket in figure~\ref{figwvltcnt}(h) has a longer temporal extent with sustained turbulence in the wake of the leading spot. 
The relative dominance of the forcing frequency in the leading region of cooler wavepacket (arrow A4) is again consistent with the destabilization of second-mode instability due to wall cooling. 
Prior high fidelity simulations \citep{krishnan2006effect, redford2012numerical} have observed a progressive dominance of second-mode waves in turbulent spots with increasing Mach numbers. 
The current analysis shows that, even under breakdown scenarios at hypersonic speeds, turbulence spots in cooler boundary layers retain a stronger presence of the initiating linear instabilities, particularly in the leading region. 

\begin{figure}
\centering
\setlength\fboxsep{0pt}
\setlength\fboxrule{0pt}
\fbox{\includegraphics[width=5.0in]{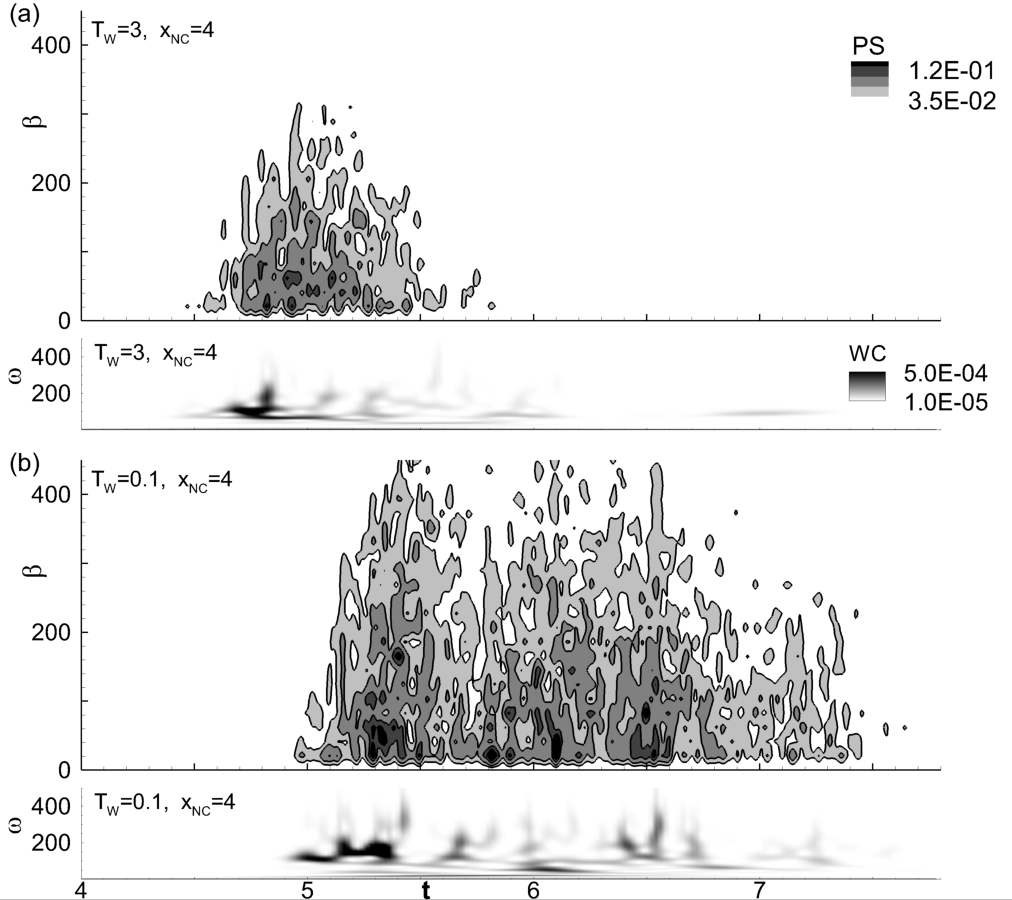}}
\caption{Spanwise wavenumber spectrum obtained as a function of time at for (a) $T_W=3$, and (b) $T_W=0.1$, near the exit of the computational domain, $X_{NC}=4$. 
In each panel, the wavenumber spectrum is augmented with the corresponding pressure scalogram on the mid-span. 
Contour levels displayed are indicated in (a).}
\label{figwvltkzext}
\end{figure}

The intermittently energized frequency spectrum trailing the head region also results in spatial scale variations in the boundary layer. 
This is evaluated using the time-accurate spanwise wavenumber distribution near the outlet of domain, $x_{NC}=4$. 
Results are provided in figure~\ref{figwvltkzext} for $T_W=3$, and $T_W=0.1$, using wall-pressure perturbations.   
The corresponding scalograms (discussed above in figures~\ref{figwvltcnt}(g) and (h)) are also reproduced again for easy reference. 
Note that in order to highlight the 3D nature of the waves, the zero wavenumber component has been removed from the spanwise signal prior to obtaining the spectrum. 
Consistent with the frequency signature, the spatial scales in the warmer-wall wavepacket are most energetic and display broadband behaviour only in the head region (figure~\ref{figwvltkzext}(a)). 
No significant energy content is evident in the oblique waves after the compact head region convects out of the domain. 
The cooler-wall wavepacket on the other hand has a longer signature on the exit plane as seen in figure~\ref{figwvltkzext}(b). 
Intermittent spurts of energy in the frequency spectrum are accompanied by the corresponding generation of 3D structures in the boundary layer. 
The presence of higher wavenumbers in the cooler-wall wavepacket confirms the presence of fine-scale features in these boundary layers.

\subsection{Characteristic features of wavepackets}\label{sec_cfwpd}
A clear understanding of formation and evolution of wavepackets is crucial to the accurate prediction and control of transition pathways in high-speed boundary layers. 
Characterization of their key features and statistics also aid in modeling their impact on the boundary layer. 
For example, measured spatio-temporal correlations were used by \citet{park2009wall} to model wall-pressure perturbations in transition zones of subsonic boundary layers. 
We now identify some important physical features of the wavepackets and turbulent spots generated in the cooled HBLs. 

\begin{figure}
\centering
\setlength\fboxsep{0pt}
\setlength\fboxrule{0pt}
\fbox{\includegraphics[width=5.0in]{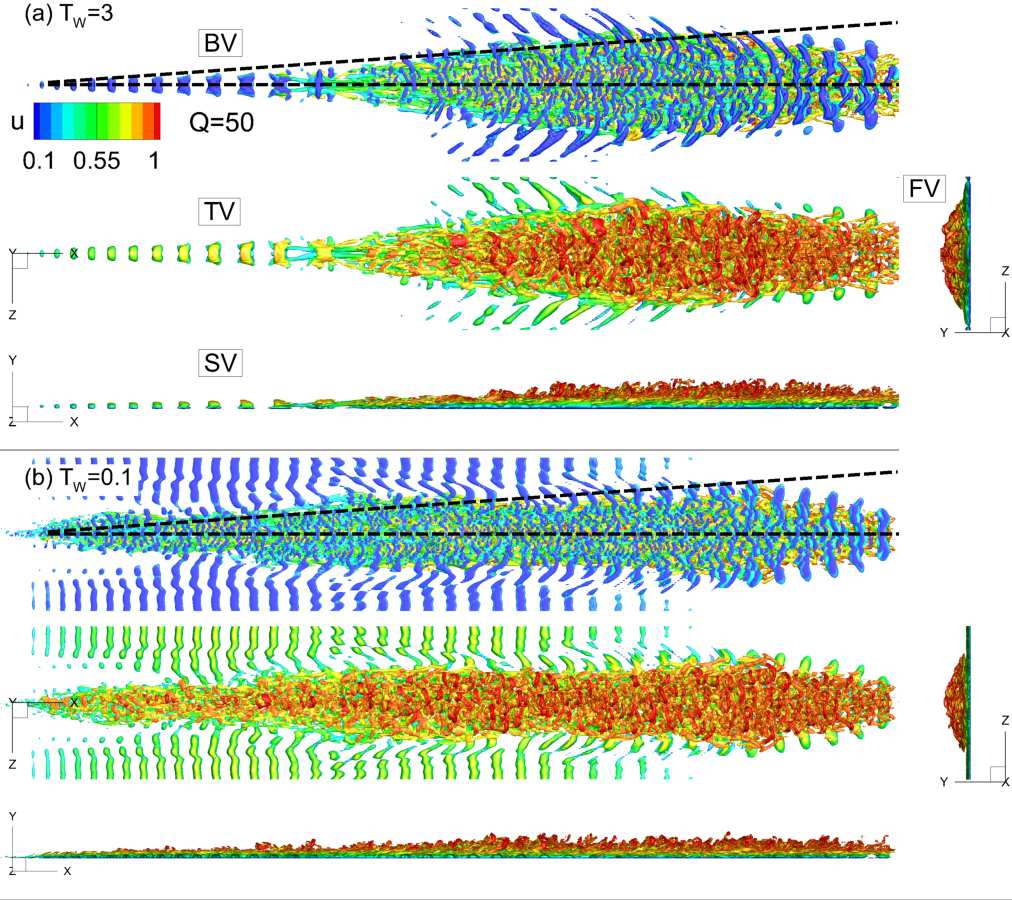}}
\caption{Physical structure of the turbulent spots generated in (a) $T_W=3$ and (b) $T_W=0.1$, as seen in  primary views, visualized using Q-criterion, colored by $u$. 
The two dashed lines in the top view of each panel mark a $4^{\circ}$ angle with the horizontal axis. 
All panels display identical contour levels, as indicated in (a). }
\label{figqtbsv}
\end{figure}

The 3D structures in the $T_W=3$ and $T_W=0.1$ wavepackets are shown in figure~\ref{figqtbsv}(a) and (b), respectively. 
Vortical features are visualized using the Q-criterion, colored by the streamwise velocity component. 
For both cases, views of the turbulent spots from beneath (BV), top (TV), side (SV) and front (FV) \textit{i.e.,} looking upstream from the outflow boundary, are provided for clarity. 
Comparing SV and FV, it is evident that the turbulent spots over warmer-walls protrude higher into the freestream, due to thicker boundary layers in the corresponding laminar profiles. 
As seen in the BV, both wavepackets contain spanwise coherent structures near the wall, which have been associated with second-mode instabilities \citep{jocksch2008growth}. 
In both wavepackets, these spanwise structures are most dominant near the leading edge of the spot, which is consistent with the presence of the second-mode frequency in the scalogram at the corresponding location (see discussion of figure~\ref{figwvltcnt}(g)). 
The lateral spreading half-angle of the turbulent spot over the warmer-wall is approximated by the dotted lines, subtending an angle of $4^{\circ}$, with the horizontal axis. 
This is consistent with experimental \citep{fischer1972spreading} and computational estimates \citep{redford2012numerical}. 
Although the cooler-wall HBL has a longer and more slender turbulent spot, the spreading rate is apparently reasonably well-approximated by the same angle. 
Considering that the wall temperature is reduced by over an order of magnitude compared to the warmer-wall, this suggests that the edge Mach number has more prominent influence, compared to wall thermal conditions, on the lateral spread rate. 
The spatial coherence of outer-boundary-layer structures is also different in the two spots, as seen from the TV images. 
The warmer-wall tends to sustain hairpin vortices that exhibit larger spanwise extent, compared to the cooler-wall, where these vortices are disintegrated. 
This is also consistent with the dominance of energy in higher $\beta$-values, observed in the spanwise spectrum of the cooler-wall HBL in figure~\ref{figwvltkzext}(b). 

A measure of the coherence in above vortical structures can be obtained from the trends in streamwise vorticity, $\omega_x$. 
These also provide information about near-wall effects of coherent structures as reported by \citet{ha2019dnsfc}, who analyzed streamwise vorticity contours and the baroclinic vorticity production term to explain the observed skin friction and heat transfer peaks over a hypersonic flared cone. 
The energetically dominant vortical structures (as represented by streamwise vorticity) in the turbulent spots formed in $T_W=3$ and $T_W=0.1$ are extracted using the proper orthogonal decomposition technique (POD) \citep{lumley1967structure}. 
\begin{figure}
\centering
\setlength\fboxsep{0pt}
\setlength\fboxrule{0pt}
\fbox{\includegraphics[width=5.0in]{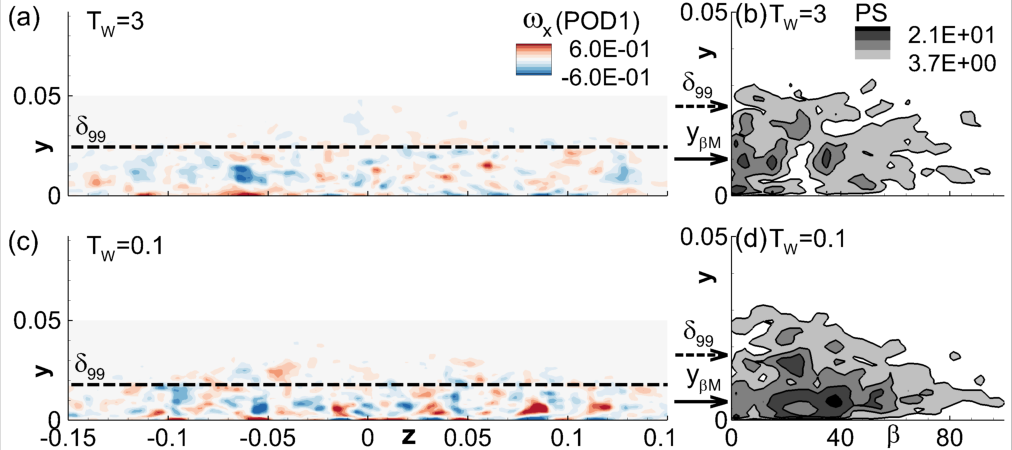}}
\caption{(a) Contours of streawise vorticity, $\omega_x$, as observed on the exit plane, $x_{NC}=4$, in the leading POD mode for $T_W=3$. 
(b) Spanwise wavenumber spectrum of the vorticity field, obtained as a function of wall-normal distance. 
(c) and (d) - corresponding results for $T_W=0.1$. 
The horizontal dashed lines in (a) and (c) mark the laminar boundary layer thickness, $\delta_{99}$. 
The arrows on the vertical axes in (b) and (d) mark $\delta_{99}$ and the location of the peak in the spanwise spectrum, $y_{\beta M}$. 
Panels (a) and (c) display identical contour levels, as indicated in (a). 
Panels (b) and (d) display identical contour levels, as indicated in (b).}
\label{figpodvrmd}
\end{figure}
The results are provided in figure~\ref{figpodvrmd}. 
POD is performed on the three-component velocity field on the exit plane defined by, $x_{NC}=4$, $0 \le y \le 0.05$, and $-0.15 \le z \le 0.15$. 
The method of snapshots \citep{sirovich1987turbulence} utilizes data on this plane during the time-interval when the turbulent region in the spot passes through.
Figures~\ref{figpodvrmd}(a) and (b) represent the contours of $\omega_x$ on the exit plane at $x_{NC}=4$, and its spanwise spectrum obtained at each wall-normal location, respectively, for $T_W=3$. 
The vorticity contours are created from the leading velocity POD mode (excluding the temporal mean), and are denoted $\omega_x(POD1)$. 
Figures~\ref{figpodvrmd}(c) and (d) are the corresponding results for $T_W=0.1$. 
While the leading mode in $T_W=3$ accounts for around $35\%$ of energy content, that in $T_W=0.1$ constitutes approximately $25\%$. 
This is due to the 
%higher rate of turbulization 
more advanced state of the breakdown
of the spot in the cooler HBL, which shifts the system away from low-rank behavior. 
The vortical structures in $T_W=0.1$ have peak energy closer to the wall ($y_{\beta M} \sim 0.05$) than those in $T_W=3$ ($y_{\beta M} \sim 0.1$), which can result in higher shear (discussed below) on the surface.  
The cooler HBL also has more instances of high-vorticity events near the wall, as indicated by the vorticity contours. 
The dominant vortical mode also contains more fine-scale features in $T_W=0.1$, which are evident by comparing the spectra in figures~\ref{figpodvrmd}(b) and (d). 
Most of the high-shear events exist below the laminar boundary layer height, $\delta_{99}$, as marked in the figures. 
This confirms the ``flattening'' of the turbulent spot in cooler HBLs, as seen earlier in the isolevels of Q-criterion in figure~\ref{figqiso4case} and figure~\ref{figqtbsv}. 

\begin{figure}
\centering
\setlength\fboxsep{0pt}
\setlength\fboxrule{0pt}
\fbox{\includegraphics[width=5.0in]{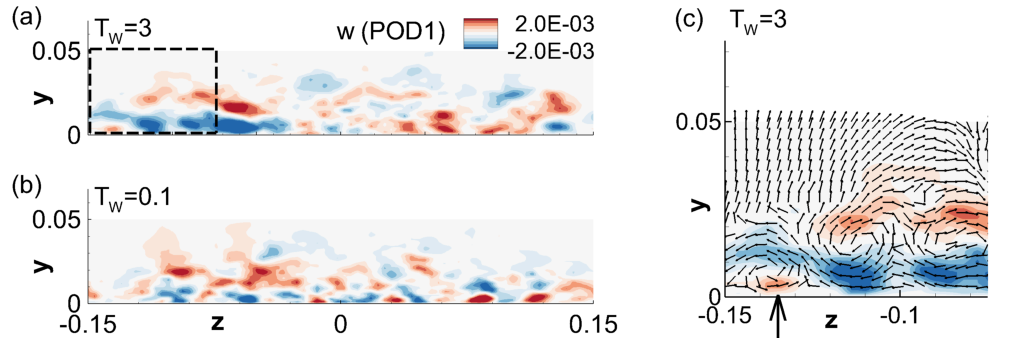}}
\caption{Contours of spanwise velocity, $w$, as observed on the exit plane, $x_{NC}=4$, in the leading POD mode for (a) $T_W=3$ and (b) $T_W=0.1$. 
The region marked by the dashed rectangle in (a) is magnified in (c). 
The vector field in (c) corresponds to $(w,v)$, as defined in the leading POD mode. 
The arrow on the horizontal axis in (c) marks a vortex beneath the lateral jet. 
All panels display identical contour levels, as indicated in (a). 
}
\label{figltjtpod}
\end{figure}

The cooler turbulent spot appears more slender and elongated in figure~\ref{figqtbsv}, compared to the warmer case. 
This could suggest variations in the lateral spreading mechanism active in these HBLs. 
\citet{redford2012numerical} identify near-wall regions of spanwise-velocity peaks, which could widen the spot by forming lateral jets. 
The effect of wall-cooling on this mechanism is analyzed in figure~\ref{figltjtpod}, by using spanwise-velocity variations. 
To ensure statistical relevance, the associated features are obtained using the leading spanwise-velocity POD mode, marked $w(POD1)$. 
Results for $T_W=3$ and $T_W=0.1$ are provided in figures~\ref{figltjtpod}(a) and (b), respectively. 
As observed in the vorticity fields above, the spanwise velocity also contains relatively larger length scales in the warmer boundary layer. 
Towards the extremities of the span ($z \sim \pm 0.15$), $T_W=3$ displays coherent regions of velocity perturbations, consistent with an outward motion from the centerline. 
Such a coherent feature is absent in the cold case. 
Enhanced turbulent breakdown of spanwise coherent structures and relatively high wavenumbers near the wall (see figure~\ref{figpodvrmd} (d)) could weaken this mechanism, thus generating a relatively slender spot. 
A detailed view of the lateral jet (marked by the dashed rectangle in figure~\ref{figltjtpod}(a)) in the warmer spot is provided in figure~\ref{figltjtpod}(c), along with the vector field, $(w,v)$, formed by its leading POD modes. 
The lateral jets originate from near the centerline of the spot and extend to its extremities. 
The upwash of low-momentum fluid from near the wall to the outer edges is evident in this field. 
Streamwise vortices exist above and beneath the lateral jet, which are accompanied by movement of fluid into the core of the spot. 
The vertical arrow marks a streamwise vortex beneath the lateral jet, resulting in the injection of near-wall flow into the turbulent spot.
This is consistent with the observations from instantaneous $\omega_x$-snapshots by \citet{redford2012numerical}, for a Mach~$3$ hot-wall boundary layer. 

\begin{figure}
\centering
\setlength\fboxsep{0pt}
\setlength\fboxrule{0pt}
\fbox{\includegraphics[width=5.0in]{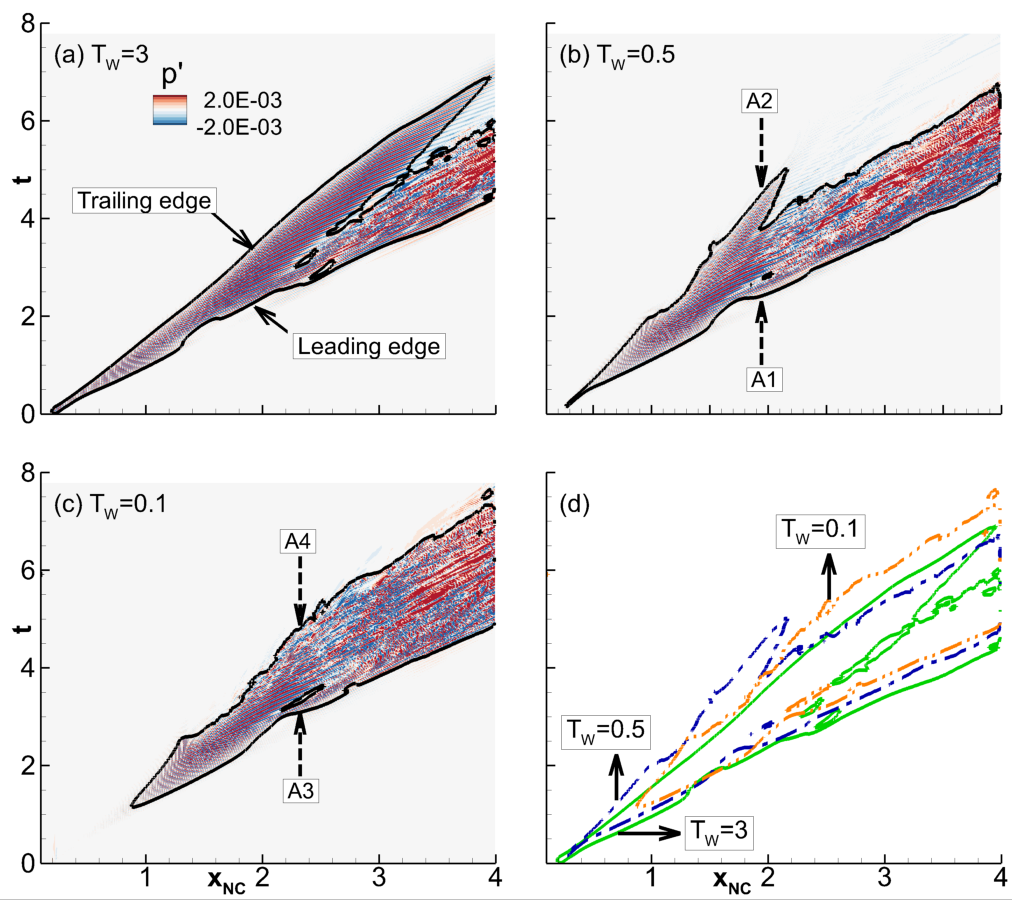}}
\caption{Chronogram of wall-pressure perturbations on the mid-span for (a) $T_W=3$, (b) $T_W=0.5$, and (c) $T_W=0.1$. 
The solid contour lines in (a), (b) and (c) trace the envelope of the evolving wavepackets. 
These three envelopes are overlaid in (d). 
Arrows A1, A2, A3 and A4 highlight relevant features in respective panels. 
Panels (a), (b) and (c) display identical contour levels, as indicated in (a).}
\label{figpxtcr}
\end{figure}

An important feature of wavepackets and the succeeding turbulent spots is their convective speed. 
Trends in this property are summarized in figure~\ref{figpxtcr}, using results from $T_W=3$, $T_W=0.5$, and $T_W=0.1$. 
The chronogram of wall-pressure perturbations along the mid-span for $T_W=3$ is shown in figure~\ref{figpxtcr}(a). 
The solid outline marks the envelope of the wavepacket at each time instant, using a predefined threshold, $|p'|=2 \times 10^{-3}$. 
Figures~\ref{figpxtcr}(b) and (c) are the corresponding results for $T_W=0.5$ and $T_W=0.1$, respectively. 
Figure~\ref{figpxtcr}(d) superimposes the envelopes from these three cases for comparison. 
The leading edges of the wavepackets have smaller slopes, $\Delta t / \Delta x_{NC}$, compared to the corresponding trailing edges, suggesting a higher convective speed in all cases. 
For $T_W=3$, the leading and trailing edges convect at approximately $85\%$ and $55\%$ of freestream velocity, respectively, which is consistent with prior experimental and computational estimates \citep{wygnanski1976turbulent,wygnanski1982spreading,jocksch2008growth}. 

Prior to addressing the effect of wall-cooling on the propagation speed, it is essential to account for the dynamical variations induced on the streamwise structure of the wavepackets.  
The outline of the envelope for $T_W=3$ indicates a clear bifurcation of the wavepacket into two zones, at $(x_{NC},t) \sim (2.5,3.8)$, which form the head and the trailing regions, earlier identified in the time-frequency analysis (see {\em e.g}, figure~\ref{figwvltcnt}(c)). 
While the head region has a chaotic imprint due to the broadband nature of its constituent frequencies and oblique modes, the trailing region is relatively well-ordered, due to its harmonic nature, and is dominated by the subharmonic of the forcing frequency. 
In cooler-wall HBLs, this trailing zone is progressively curtailed, as seen from the trends in figures~\ref{figpxtcr}(a), (b) and (c). 
The streaks in the turbulent head-region emerge as the major component of the wavepacket trace in the cooler-walls, and exhibit a different convective speed than the precursor harmonic wavepacket. 
This switch in the convective speeds is highlighted by two vertical dashed arrows; A1 and A2 in figure~\ref{figpxtcr}(b), and A3 and A4 in figure~\ref{figpxtcr}(c). 
The convective speeds of the leading edges increase by around $10\%$ in the cooler-walls after 
%turbulization. 
initiation of breakdown.
The trailing edges exhibit a more dramatic increase, by over $60\%$.

\subsection{Wall loading}\label{sec_wlld}
\begin{figure}
\centering
\setlength\fboxsep{0pt}
\setlength\fboxrule{0pt}
\fbox{\includegraphics[width=5.0in]{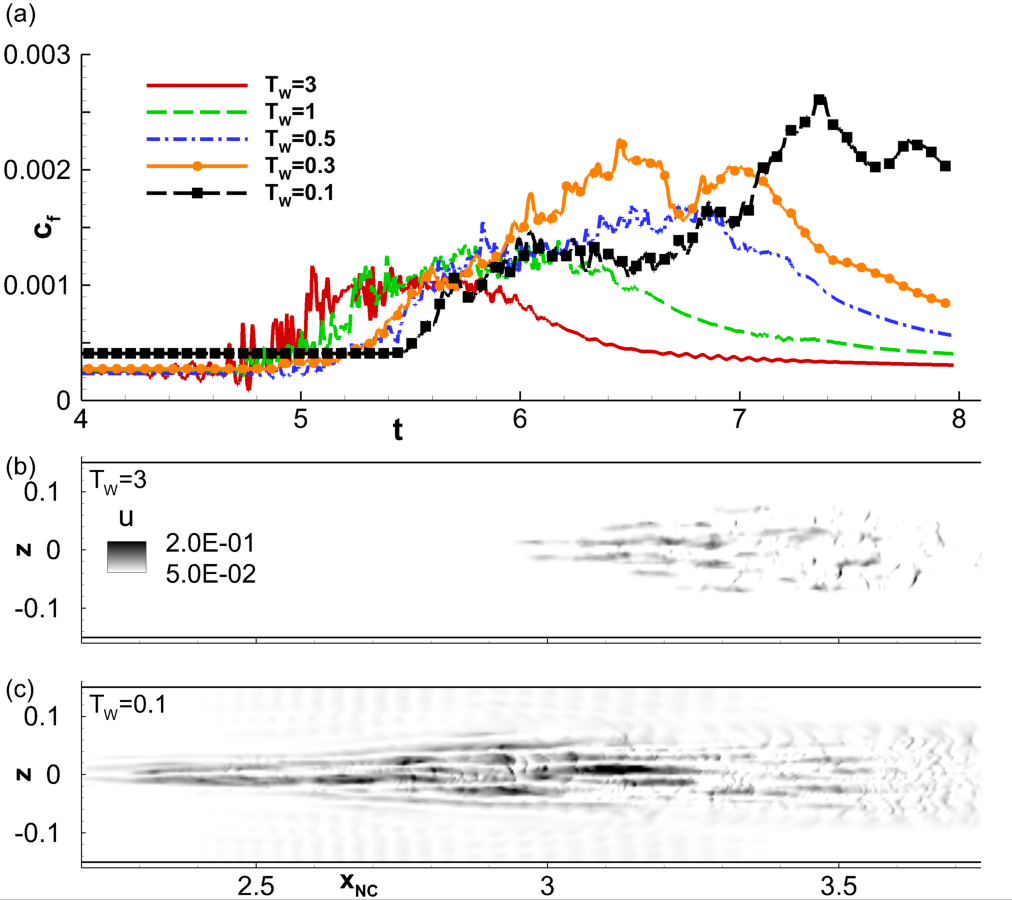}}
\caption{(a) Temporal evolution of spanwise averaged skin friction coefficient, $c_f$, at $x_{NC} =4$, for various cold walls. 
Instantaneous near-wall contours of streamwise velocity for (b) $T_W=3$ and (c) $T_W=0.1$. 
Panels (b) and (c) display identical contour levels, as indicated in (b).
}
\label{figsknfrc}
\end{figure}

Wavepackets that cause transition in HBLs result in increased drag and thermal loads on the surface. 
The impact on drag due to wall cooling is assessed in figure~\ref{figsknfrc}. 
Since the qualitative aspects of the wavepacket evolve rapidly, the results are visualized in a time-accurate manner near the exit-plane of the computational domain, at $x_{NC} =4$, in figure~\ref{figsknfrc}(a). 
The quantity plotted is the coefficient of skin friction, $c_f$, defined as follows:
\begin{equation}\label{cfeqs}
c_f=\frac{2}{Re} \mu \left. \frac{\partial u}{\partial y} \right |_{y=0}.
\end{equation} 
The coefficient is averaged in the spanwise direction as well. 
To summarize the trends, data from all the HBLs that transitioned are included here. 
$c_f$ increases from the laminar values as the turbulent spot approaches the exit plane, \textit{e.g.}, $t \sim 4.5$ in the case, $T_W=3$. 
This is also the location where the leading edge envelope intersects $x_{NC}=4$ in figure~\ref{figpxtcr}(a). 
For the cooler-walls, the $c_f$ increase begins at later time instants, due to the previously observed slower convective speeds of the respective wavepackets. 
The peak $c_f$ values at $T_W=3$ are consistent with those observed for fully turbulent HBLs at adiabatic conditions \citep{egorov2016direct}. 
With cooling, the values of $c_f$ increase monotonically, and peak values more than double for the highly cooled cases, $T_W=0.3$ and $T_W=0.1$. 
The temporal distribution of the $c_f$-peaks also changes qualitatively. 
For $T_W=3$, the maximum is localized near the leading edge of the spot where dominant hairpin vortices exist. 
Behind this region, near-wall shear weakens, and is reflected as low values of $c_f$. 
In the cooler-walls, the spots are slender and elongated, thus resulting in a broader hump in $c_f$. 
In addition, multiple local peaks are observed due to the intermittent generation of turbulent zones in the wake of the leading spot.
Thus, wakes of spots with sustained generation of turbulence have greater impact on wall loading than the spatially localized leading spot observed in the warmer-walls. 

The distributed impact of cooler wavepackets on the wall is confirmed in figures~\ref{figsknfrc}(b) and (c), through the near-wall streamwise velocity contours. 
These velocity fields correspond to the time-instant at which vortical features are displayed in figure~\ref{figqtbsv}.
Compared to $T_W=3$ (figure~\ref{figsknfrc}(b)), the high-speed streaks in $T_W=0.1$ (figure~\ref{figsknfrc}(c)) are elongated, and peaks values appear significantly downstream of the leading edge. 
The higher velocities present in these streaks are evident in the contour levels, and result in increased drag on the surface. 

\begin{figure}
\centering
\setlength\fboxsep{0pt}
\setlength\fboxrule{0pt}
\fbox{\includegraphics[width=5.0in]{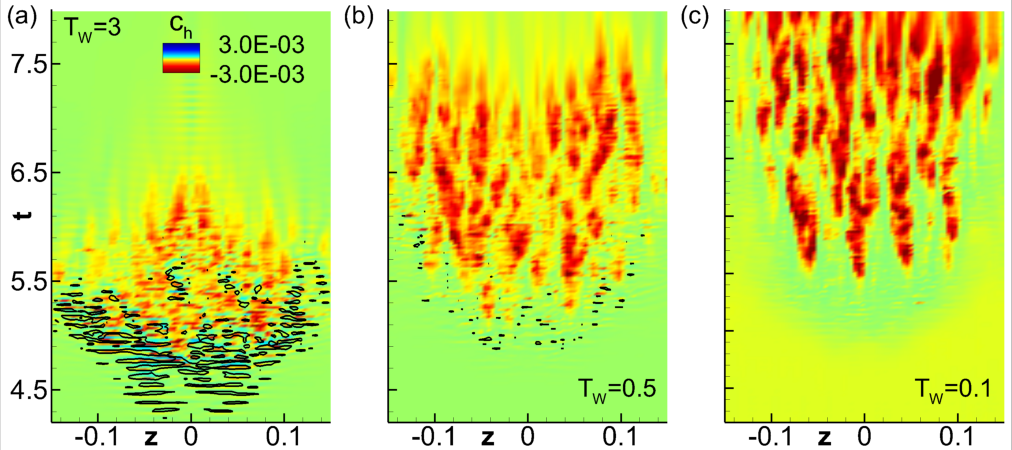}}
\caption{Temporal evolution of surface heat transfer coefficient, $c_h$, at $x_{NC} =4$, for (a) $T_W=3$, (b) $T_W=0.5$ and (c) $T_W=0.1$. 
The solid contour lines in (a) and (b) mark $c_f=0$. 
All panels display identical contour levels, as indicated in (a).}
\label{figsrfdtdy}
\end{figure}

The effects on wall heat transfer are reported in figure~\ref{figsrfdtdy}, which plots the surface heat transfer coefficient, $c_h$, at $x_{NC} =4$. 
$c_h$ is defined as follows \citep{novikov2017direct}:
\begin{equation}\label{cheqs}
c_h=\frac{Q_W^{*}}{\rho_\infty^{*}U_\infty^{*}c_p^{*}(T_r^{*} - T_W^{*})}=-\frac{\mu_w}{Re Pr} \frac{1}{(T_r - T_W)} \left. \frac{\partial T}{\partial y} \right |_{y=0}.
\end{equation} 
$Q_W^{*}=-\mathcal{K}^{*} \left. \frac{\partial T^{*}}{\partial y^{*}} \right |_{y^{*}=0}$ is the wall heat transfer rate, and $\mathcal{K}^{*}$ is the thermal conductivity. 
Recovery temperature, $T_r^{*}$ is calculated as:
\begin{equation}\label{rctmpeqs}
\frac{T_r^{*}}{T_\infty^{*}} = 
1 + r \frac{\gamma -1}{2}M_\infty^{2}.
\end{equation} 
Since the plots are obtained near the exit plane of the breakdown region of the spots, the recovery factor, $r=Pr^{1/3}$.

Typical $c_h$ values in the streaks formed beneath the spot for $T_W=3$ are about $-2 \times 10^{-3}$, consistent with turbulent estimates in HBLs \citep{chynoweth2019history}. 
This increases to approximately $-3 \times 10^{-3}$ and $-4 \times 10^{-3}$, for $T_W=0.5$ and $T_W=0.1$, respectively. 
Note that the negative values indicate the presence of relatively warmer fluid near the surface, resulting in heat transfer to the surface. 
In supersonic boundary layers at near-adiabatic-wall temperatures, \citet{krishnan2006effect} report positive values of $c_h$ (wall cooling) in the calmed region behind the turbulent spot. 
In order to identify such regions in the HBLs under consideration, the $c_h$ flood-contours are overlaid with the $c_h=0$ contour using solid lines. 
For $T_W=3$ (figure~\ref{figsrfdtdy}(a)), these zero-contour lines identify regions of wall-cooling along the leading regions of the turbulent spot. 
The ``inrush'' \citep{krishnan2006effect} of relatively warm fluid in the trailing region of these spots results in wall-heating, as opposed to cooling in near-adiabatic cases. 
As the wall is increasingly cooled, the regions of zero-contours diminish as seen for $T_W=0.5$ and $T_W=0.1$, in figures~\ref{figsrfdtdy}(b) and (c), respectively. 
In addition, the trailing streaks become more elongated with higher temperature gradients, resulting in sustained and elevated rates of heat transfer to the surface. 
Thus, although the leading region of cold wall wavepackets are energetically dominant (see figure~\ref{figwvltcnt}(h)), peak wall loading occurs in the elongated trailing region. 
Since the extent of the trailing region is highly dependent on its convective velocity, modeling efforts should accurately account for the variations in its value, as identified in figure~\ref{figpxtcr}.

\section{Summary}\label{sec_cln}
Variations in linear instability mechanisms and flow features due to wall-cooling in a hypersonic boundary layer (HBL) at Mach~6 are studied, with particular emphasis on the impact on transition. 
The dynamics of instability waves associated with wall cooling confirm that over sufficiently cooled walls, the unstable mode~F, after synchronizing with the slow continuous acoustic spectrum, switches from a compact form to an oscillatory eigenfunction; this results in supersonic phase speeds in the outer boundary layer and radiation into the freestream. 
A physical analysis of this supersonic mode using momentum potential theory (MPT) indicates that radiation from cooled HBLs contains similar acoustic and vortical wave magnitudes. 
The entropic component is however, negligible. 

Perturbation analysis of the mean flow associated with  varying degrees of wall cooling, using a high-order numerical approach, confirms the observations from linear theory and adds further insight. 
While the subsonic second-mode displays a near-constant wavenumber distribution in the HBL, that in the supersonic mode gradually increases in the outer boundary layer, as it aligns towards the wall. 
In moderately cooled walls, the second-mode retains its trapped nature between the wall and the critical layer, as seen previously in near-adiabatic walls. 
At higher levels of cooling, however, this trapped form ruptures, resulting in efflux of energy into the freestream. 
%This channels a portion of fluctuating energy from the HBLs, which is quantified by calculating the vortical and acoustic efflux of total fluctuating enthalpy. 
This efflux is quantified by calculating the vortical and acoustic components of the total fluctuating enthalpy. 
Although wall-cooling increases the radiative loss, there is no apparent reduction of perturbation growth in the HBLs. 
Increased wall cooling monotonically increases wall-pressure perturbations, in line with the destabilization of second-mode. 

The role of supersonic modes, and mean flow variations due to wall cooling, in inducing realistic transition scenarios are addressed through the study of three-dimensional (3D) wavepacket-evolution.
%in moderately to highly cooled HBLs. 
The choice of forcing highlights the effect of wall-cooling on transition; for the near-adiabatic wall, the second-mode wavepacket of the chosen attributes attenuates after an initial growth region.
For cooled HBLs however, the same forcing induces wavepackets that nonlinearly saturate and generate turbulent spots, despite the observed radiative losses. 
Following the percolation of energy into the subharmonic of forcing frequency, all wavepackets elongate in the streamwise direction, and display a harmonic nature. 
In moderately cooled walls, the next phase involves bifurcation of the wavepacket into a dominant head region and a harmonic trailing region. 
The head region contains a broadband spectrum, and increasingly displays breakdown of the HBL with wall cooling, eventually resulting in a localized turbulent spot. 
The highly cooled walls exhibit a qualitatively different scenario, where the wavepackets are significantly more elongated in the streamwise direction. 
The head region, although energetically dominated by the second-mode forcing frequency, displays a prominent wake, where intermittent local spurts of turbulence are observed. 
This leads to a broader range of spatio-temporal scales in the cooler-wall wavepacket, which are quantified through spectral analyses. 

Key physical features of the wavepackets including convective speeds and wall loading are characterized, which can aid in transition model development in cold wall conditions.
The moderately cooled wavepackets conform to existing estimates of leading and trailing edge velocities. 
Once turbulent regions emerge in the highly cooled cases, the convective speed of the trailing edge increases dramatically by over $60\%$. 
The wakes in the highly cooled wavepackets result in significantly elongated near-wall high velocity streaks. 
These induce sustained values of peak skin friction, that are about twice those reported for adiabatic walls. 
Heat transfer rates also more than double in the highly cooled walls, and primary regions of wall heating exist in the trailing region of the turbulent spots.  
Thus, accurate estimates of convective speeds of turbulent spots are critical to reliable wall loading estimates in highly cooled HBLs.
%??? SEE COMMENT

\section*{Declaration of Interests}
Declaration of Interests. The authors report no conflict of interest

\section*{Acknowledgments}
This research was supported by the Office of Naval Research (Grant: N00014-17-1-2528) monitored by E. Marineau, with R. Burnes serving as technical point of contact.
The simulations were performed with a grant of computer time from the DoD HPCMP DSRCs at AFRL, NAVO and ERDC, and the Ohio Supercomputer Center.

\section*{References}
\bibliography{MASTER_CHBL}

\end{document}